\documentclass[reqno,final]{amsart} 

\usepackage{amssymb}
\usepackage[T1]{fontenc}
\usepackage[british]{babel}
\usepackage{graphicx}
\usepackage[utf8]{inputenc}
\usepackage{mathrsfs}
\usepackage{subfigure}
\usepackage{xspace}
\usepackage{enumerate}
\usepackage[initials]{amsrefs}
\usepackage[pdfpagelabels,pdftex]{hyperref}
\hypersetup{%
        linkcolor=black,                
        urlcolor=blue,                 
        breaklinks=true,                
        colorlinks=false,               
        citebordercolor=0 0 0,          
        filebordercolor=0 0 0,          
        linkbordercolor=0 0 0,          
        menubordercolor=0 0 0,          %
        urlbordercolor=0 0 0,           
        pdfhighlight=/I,                %
        pdfborder=0 0 0,                
}

\newcommand{\R}{\ensuremath{\mathbb R}\xspace}
\newcommand{\C}{\mathbb C}
\newcommand{\St}{\ensuremath{\mathbb{S}^2}\xspace}
\newcommand{\Sth}{\ensuremath{\mathbb{S}^3}\xspace}
\newcommand{\U}{\ensuremath{\mathrm{U(1)}}\xspace}
\newcommand{\SU}{\ensuremath{SU(2)}\xspace}
\newcommand{\mbar}{\overline{m}}

\newcommand{\lie}[2]{\mathcal L_{#1}{#2}}
\newcommand{\etp} {\eth'}

\newcommand{\ethmath}{\eth} 
\newcommand{\commutator}[3]{{C^#1}_{#2 #3}}
\newcommand{\Eqref}[1]{Eq.~(\ref{#1})}

\newcommand{\Sectionref}[1]{Section~\ref{#1}}
\newcommand{\pairing}[2]{\left<#1,#2\right>}
\newcommand{\highlight}[1]{\emph{#1}\xspace}
\renewcommand*{\i}{\mathrm{i}}

\title[A spectral method based on spin-weighted spherical harmonics]{Numerical evolutions of fields on the 2-sphere using a spectral method based on spin-weighted spherical harmonics}

\author[F. Beyer]{Florian Beyer}
\address{Department of Mathematics and Statistics, University of Otago, PO Box 56, Dunedin 9010, New Zealand}
\email{fbeyer@maths.otago.ac.nz}
\author[B. Daszuta]{Boris Daszuta}
\address{Department of Mathematics and Statistics, University of Otago, PO Box 56, Dunedin 9010, New Zealand}
\email{dasbo917@student.otago.ac.nz}
\author[J. Frauendiener]{J\"org Frauendiener}
\address{Department of Mathematics and Statistics, University of Otago, PO Box 56, Dunedin 9010, New Zealand}
\email{joergf@maths.otago.ac.nz}
\author[B. Whale]{Ben Whale}
\address{Department of Mathematics and Statistics, University of Otago, PO Box 56, Dunedin 9010, New Zealand}
\email{bwhale@maths.otago.ac.nz}

\numberwithin{equation}{section}

\begin{document}
\begin{abstract}
 Many applications in science call for the numerical simulation of systems on manifolds with spherical topology.  Through use of integer spin weighted spherical harmonics we present a method which allows for the implementation of arbitrary tensorial evolution equations.  Our method combines two numerical techniques that were originally developed with different applications in mind.  The first is Huffenberger and Wandelt's spectral decomposition algorithm to perform the mapping from physical to spectral space.  The second is the application of Luscombe and Luban's method, to convert numerically divergent linear recursions into stable nonlinear recursions, to the calculation of reduced Wigner d-functions.  We give a detailed discussion of the theory and numerical implementation of our algorithm.  The properties of our method are investigated by solving the scalar and vectorial advection equation on the sphere, as well as the $2+1$ Maxwell equations on a deformed sphere.
\end{abstract}
\maketitle

\section{Introduction}

We present a method for the numerical calculation of solutions to general hyperbolic partial differential equations (PDEs) over the sphere that combines several existing techniques in novel ways.  Our motivating interest is the analysis of gravitational radiation in a neighbourhood of infinity via Friedrich's conformal field equations \cite{friedrich98:_gravit_fields}, where it is convenient to regard space-time as the product of a $2$ dimensional Lorentzian manifold and a Euclidean $2$-sphere.  The main difficulty for the numerical solution of problems in this geometric setting stems from the fact that $\mathbb{S}^2$ cannot be covered by a single chart.  Hence the coordinate description of fields inevitably breaks down.  Spectral methods avoid most of the problems caused by this as fields on $\mathbb{S}^2$ are expressed as a sum of functions that form a well defined bases on the sphere. Assuming that the properties of the functions in the sum are well understood, it is possible to avoid working directly with coordinate expressions that suffer from coordinate singularities. As an example, we mention the recent work of \cite{Csizmadia:2013multi,Racz:2011tails} in the context of Kerr space-time shows that indeed issues such as coordinate singularities and instability are avoided in such an approach, leading to accurate evolutions of dynamical equations. As a consequence we use a spectral method in what follows.

To implement our spectral method we choose to work with spin weighted spherical harmonics (SWSHs) and the associated $\eth$ and $\eth'$ operators \cite{newman_note_1966,Penrose:1984tf}.  SWSHs are a generalisation of scalar, vector and tensor harmonics on the sphere \cite{thorne_multipole_1980,SPHEREPACK}.  SWSHs form an orthonormal basis for $L^2(SU(2))$ \cite{Goldberg:1967vm} upon which the differential operators $\eth$ and $\eth'$ act by raising and lowering spin-weight, respectively.  In common with all spectral methods, this reduces the action of differential operators to algebraic manipulations.  This is a property which we exploit to reduce PDEs to systems of coupled ODEs.

In order to apply SWSHs to a spectral evolution scheme 
it is necessary to
decompose arbitrary fields on $\mathbb{S}^2$ into a sum of SWSHs. 
Many methods to do this have been proposed, since such decompositions
are important for the analysis of data over the sphere, e.g.\
\cites{Doroshkevich:2005GLpxlCmb,Gorski:2005healpix,Kostelec:2008fftRotGrp,
McEwen:2011:NovelSampling}.
We choose 
Huffenberger and Wandelt's method \cite{Huffenberger:2010hh} 
(which is a modification of \cite{Dilts:1985SphHarmExpa}
and \cite{McEwen:2011:FaExSpinSph})
for three
reasons. Firstly, the algorithm is theoretically exact
if a minimum number of grid points are used. This is
in contrast to several alternatives
that are asymptotically exact
in the limit of increasing numbers of grid points
\cite{Alpert:1991:LegendreExp,Suda:2002Fastsph,Rokhlin:2004Fastsph}. 
Secondly, the method can be applied
simultaneously to functions of different spin weights. In our
desired application we will be working with several functions each with
different spin weights \cite{Beyer:2012ie}. 
As a consequence Huffenberger and Wandelt's
method has reduced computational effort in comparison to
a method which operates on each spin weighted function separately.
Thirdly, Huffenberger and Wandelt's method, via a clever mapping
of the sphere into the $2$-torus allows for Fast Fourier Transformations
to be used, see also
\cites{bartnik_einstein_1999,Beyer:2009vw,Muciaccia:2009fastSphHarm}. 
This is in contrast to
spectral 
methods adapted to the non-periodic coordinate on $\mathbb{S}^2$,
e.g.\  \cite{Doroshkevich:2005GLpxlCmb}.

In order to calculate the values of SWSHs over $\mathbb{S}^2$ Huffenberger and Wandelt used a formula relating SWSHs to reduced Wigner $d$-functions evaluated at $\frac{\pi}{2}$.  They implemented the calculation of these $d$-functions via the three term linear recursive relations given by Risbo and Trapani and Navaza \cite{Trapani:2006he,Risbo:1996iy}.  Trapani and Navaza's scheme is both faster and more accurate than Risbo's but eventually becomes unstable.  To cope with this use a non-linear scheme that is equivalent to the scheme proposed by Pr\'ezeau and Reinecke \cite{Prezeau:2010algoWig}.  We build a hybrid linear/non-linear recursion that avoids the numerical problems that both linear and non-linear recursive schemes suffer from, see Sections \ref{sec:algodelCalc} and \ref{sec:wdelstab}.  A similar hybrid scheme has been proposed by Luscombe and Luban \cite{Luscombe:1998:SimplifiedW3jNL} for the calculation of $3j$ and $6j$ symbols. To the best of the authors' knowledge neither the use of a hybrid scheme for the calculation of reduced Wigner $d$-functions at $\frac{\pi}{2}$ nor the use of this method in SWSH decompositions of functions over $\mathbb{S}^2$ has been used before.

The $2+1$ Maxwell equations on a deformed
sphere, Section \ref{sec:maxCnstrct}, 
have non-constant coefficients.
We will therefore need to 
perform a SWSH decomposition of products of SWSHs. This requires
the calculation of Clebsch-Gordan
coefficients. 
To do this we use a hybrid linear/non-linear
recursion for $3j$ symbols originally presented by Luscome and 
Luban's method 
\cite{Luscombe:1998:SimplifiedW3jNL}
to 
Schulten and Gordan's linear
scheme \cite{Schulten:1975:exactW3j}. As before, to the best of the
author's knowledge this is the first time a hybrid linear/non-linear
scheme has been used for the calculation of Clebsch-Gordan
coefficients. An alternative to the above explicit decomposition is to use
spectral transformations directly to perform decompositions of
products of SWSHs
\cites{brugmann_pseudospectral_2013,bartnik_einstein_1999} (the
pseudo-spectral approach). We use
this approach to check for accuracy.

It is our goal to provide a self-contained presentation, with consistent conventions, that may be readily adapted to general non-linear hyperbolic PDEs using the outlined spectral method. Wherever possible we present results for both integer and half-integer spin.  This will provide a foundation for future extension to the half-integer spin case.

This paper is structured as follows: in \S\ref{sec:geom} we provide the geometric background appropriate for formulation of problems later in the work, in particular we review the construction of frames adapted for use with the $\eth$-formalism. In \S\ref{sec:weightedquantities} we explicitly show how an arbitrary smooth tensor field may be represented in terms of spin-weighted quantities on $\mathbb{S}^2$. In \S\ref{sec:introCGw3j} we introduce and review the properties of SWSHs.  We describe in detail how products of SWSHs may be decomposed and give details about the symmetry properties of Clebsch-Gordan coefficients (Wigner $3j$-symbols) appropriate for numerical use.  In \S\ref{sec:specXform} we briefly describe the spherical harmonic transformation of \cite{Huffenberger:2010hh}. In \S\ref{sec:CGw3jNumProc} we describe how we compute Clebsch-Gordan coefficients (Wigner $3j$-symbols) numerically.  In \S\ref{sec:errorStabilitySpeed} we demonstrate the property of spectral convergence for smooth test functions. In \S\ref{sec:wdelstab} we demonstrate the instability of the recursive calculation of $d^l_{mn}$ due to \cite{Trapani:2006he} and discuss how our hybrid scheme avoids this.  In \S\ref{sec:PSandFScmp} we contrast the pseudo-spectral and spectral approaches to the decomposition of products of SWSHs.  In \S\ref{sec:advProb} we construct the tensor advection equation in the $\eth$ formalism, thus showing how the standard IVP for the scalar and vector advection equation on $\mathbb{S}^2$ may be formulated; in \S\ref{sec:advNum} we numerically solve this problem using our spectral method for test fields that lead to temporally periodic solutions. Exploiting this periodicity by comparing solutions at integer multiples of one period (i.e. stroboscopically) yields a method for performing convergence tests. In \S\ref{sec:maxCnstrct} we construct the IVP for the $2+1$ Maxwell equation where the spatial geometry is conformally related to $\mathbb{S}^2$. In \S\ref{sec:numMax} we numerically solve the problem, comparing spectral with pseudo-spectral methods.


\section{Geometric preliminaries}
\label{sec:geom}

\subsection{The \texorpdfstring{$2$}{2}- and \texorpdfstring{$3$}{3}-spheres and the Hopf bundle}

It is often useful to think of the manifold \Sth as the submanifold of $\R^4$ given by $x_1^2+x_2^2+x_3^2+x_4^2=1$. The \highlight{Euler coordinates} of $\Sth$ can then be represented by
  \begin{equation*}    
    \begin{split}
      x_1&=\cos\frac\theta 2\cos\lambda_1, 
      \quad x_2=\cos\frac\theta 2\sin\lambda_1,\\
      x_3&=\sin\frac\theta 2\cos\lambda_2, 
      \quad x_4=\sin\frac\theta 2\sin\lambda_2,
    \end{split}
  \end{equation*}
  where $\theta\in (0,\pi)$ and $\lambda_1,\lambda_2\in (0,2\pi)$.
  Clearly, these coordinates break down at $\theta=0$ and $\pi$.  For later convenience, we
  introduce coordinates $(\theta,{\rho},{\phi})$ (which are also referred to as Euler coordinates) by
  \begin{equation*}
    \lambda_1=({\rho}+{\phi})/2,\quad
    \lambda_2=({\rho}-{\phi})/2.
  \end{equation*}
The set of complex unitary $2\times 2$-matrices with unit determinant $\SU$ endowed with the natural smooth manifold structure is diffeomorphic to $\Sth$
\begin{equation*}
\Psi:\Sth\rightarrow\SU,\quad (x_1,x_2,x_3,x_4)\mapsto 
\begin{pmatrix}
  x_3+\i x_4 & -x_1+\i x_2\\
  x_1+\i x_2 & x_3-\i x_4
\end{pmatrix}.
\end{equation*}
Since the latter is a Lie group, we can use the map $\Psi$ to endow $\Sth$ with a Lie group structure. In the following, it is therefore not always necessary to distinguish between $\Sth$ with $\SU$.

Any Lie group is parallelizable, and a smooth global frame on $\SU$ can be constructed as follows.  We define left and right translation maps
\[L, R: \SU\times\SU\rightarrow\SU,\quad (u,v)\mapsto L_u(v):=uv,
\quad (u,v)\mapsto R_u(v):=vu.
\] 
On any Lie group, the maps $L_u$ and $R_u$ are automorphisms for each element $u$.  Now, choose a basis of the tangent space at the unit element $T_e(\SU)$ (i.e.\ a basis of the Lie algebra) 
\begin{equation*}
  \tilde Y_1=\frac 12\begin{pmatrix}0&\i\\\i&0\end{pmatrix},\quad
  \tilde Y_2=\frac 12\begin{pmatrix}0&-1\\1&0\end{pmatrix},\quad
  \tilde Y_3=\frac 12\begin{pmatrix}\i&0\\0&-\i\end{pmatrix},
\end{equation*}
and define, for $u\in\SU$,
\begin{equation*}
Y_k: u\mapsto (Y_k)_u:=(L_u)_*(\tilde Y_k),\quad Z_k: u\mapsto (Z_k)_u:=(R_u)_*(\tilde Y_k).
\end{equation*}
Clearly, $(Y_k)=(Y_1, Y_2, Y_3)$ is a smooth global frame on \Sth which is invariant under left translations while the frame $(Z_k)$ is invariant under right translations.
These fields have the following representation with respect to the Euler parametrization 
\begin{align*}
  Y_1&=-\sin \rho\,\partial_\theta
  -\cos \rho
  \left(\cot\theta\partial_{\rho}
    -\csc\theta\partial_{\phi}\right),\\
  Y_2&=-\cos \rho\,\partial_\theta
  +\sin \rho
  \left(\cot\theta\partial_{\rho}
    -\csc\theta\partial_{\phi}\right),\\  
  Y_3&=\partial_{\rho},\\
  Z_1&=-\sin \phi\,\partial_\theta
  +\cos \phi
  \left(\csc\theta\partial_{\rho}
    -\cot\theta\partial_{\phi}\right),\\
  Z_2&=-\cos \phi\,\partial_\theta
  -\sin \phi
  \left(\csc\theta\partial_{\rho}
    -\cot\theta\partial_{\phi}\right),\\
  Z_3&=-\partial_\phi.
\end{align*}
One can show by direct Lie group arguments (or by using the coordinate representations of the fields) that
\[
[Y_1,Y_2]=Y_3,\quad[Y_2,Y_3]=Y_1,\quad[Y_3,Y_1]=Y_2,
\] 
similarly for the right-invariant fields, and
\[
[Y_k,Z_l]=0,\quad\forall k,l=1,2,3.
\]
For later convenience, we define 
\begin{equation*}
  \ethmath:=-({Y_2} + \i{Y_1}),\quad
  \ethmath':=-({Y_2} - \i{Y_1}),
\end{equation*}
which, as we shall see later, are closely related to the $\eth$-operators defined in~\cite{Penrose:1984tf}. We have
\begin{equation}
\label{eq:ethbracket}
[\ethmath,\ethmath']=2 \i Y_3.
\end{equation}

The \highlight{Hopf map} $\pi:\Sth\rightarrow\St$ can be represented as
\begin{align*}
  (x_1,x_2,x_3,x_4)\mapsto
  (y_1,y_2,y_3)
  &=(2(x_1 x_3+x_2x_4),2(x_2x_3-x_1x_4),
  x_1^2+x_2^2-x_3^2-x_4^2)\\
  &=(\sin\theta\cos{\phi},\sin\theta\sin{\phi},\cos\theta).
\end{align*}
Here we again consider \Sth as being embedded into $\R^4$ by
$x_1^2+x_2^2+x_3^2+x_4^2=1$, and the manifold \St is thought of being
given by $y_1^2+y_2^2+y_3^2=1$ in $\R^3$. When we introduce standard
polar coordinates on \St, namely
\[y_1=\sin\vartheta\cos{\varphi},\quad
y_2=\sin\vartheta\sin{\varphi},\quad
y_3=\cos\vartheta,
\]
then $\pi$ obtains the particularly simple representation
\begin{equation}
  \label{eq:localHopf}
  \pi:(\theta,{\rho},{\phi})\mapsto (\vartheta,\varphi)=(\theta,{\phi}).
\end{equation}
In particular, it becomes obvious that the push-forward of $Y_3$ to \St along $\pi$
vanishes. Indeed,  $\Sth$ is the principal bundle over
$\St$ with structure group $\U$ generated by $Y_3$ (whose integral
curves are closed) and projection map $\pi$; this is the \highlight{Hopf bundle}.

The Hopf bundle can be identified with the bundle of orthonormal frames on $\St$ with respect to any smooth Riemannian metric. An explicit construction in terms of the coordinates above can be done as follows.  Let $U$ be an open subset of $\St$; we assume that the poles $\vartheta=0,\pi$ are outside of $U$ so that the representation of the Hopf map given by \Eqref{eq:localHopf} is well-defined and the Euler coordinates cover $\pi^{-1}(U)$. If we restrict to sufficiently small open subsets this is no loss of generality since for any sufficiently small choice of the open set $U$ we can always introduce the coordinates such that the poles are not in $U$.  Let $(e_1^*, e_2^*)$ be a smooth orthonormal frame on $U$ and define the corresponding complex field 
\[
m^*:=\frac 1{\sqrt 2}(e_1^* + \i e_2^*).  
\] 
We consider the action 
\begin{equation}
  \label{eq:mmStar}
  m^*\mapsto m=e^{ \i\rho}m^*,
\end{equation}
of \U which is defined pointwise on $U$, i.e.\ the group parameter $\rho$ is a smooth function on $U$. Any specification of the function $\rho(\vartheta,\varphi)$ therefore yields another smooth orthonormal frame on $U$ and can hence be interpreted as the smooth local section $U\rightarrow\Sth, (\vartheta,\varphi)\mapsto (\theta,\phi,\rho)=(\vartheta,\varphi,\rho(\vartheta,\varphi))$ in the bundle of orthonormal frames, or equivalently, in the Hopf bundle. Doing this for every open subset $U$ of $\St$ (introducing coordinates so that the poles are not in $U$ as above), the full bundle of orthonormal frames can be recovered and can therefore be identified with the Hopf bundle. At every point $p$ of $U$, the fibre $\pi^{-1}(p)$ is the set of all orthonormal bases of $T_p(\St)$.

\subsection{Weighted quantities on the \texorpdfstring{$2$}{2}-sphere}
\label{sec:weightedquantities}

Let $(\omega,\bar \omega)$ be the dual coframe of $(m,\bar m)$ and $(\omega^*,\bar \omega^*)$ dual to $(m^*,\bar m^*)$. Then, the above action of $\U$ implies
\[
\omega=e^{-\i\rho}\omega^*.
\] 
Now let an arbitrary smooth tensor field $T$ of type $(s_1+s_2,r_1+r_2)$ for integers
$r_1,s_1,r_2,s_2\ge 0$ be given on $U$, so that the function
$\nu: U\rightarrow\C$ is defined by
\begin{equation*}
\nu:=T(
\underbrace{\omega,\ldots,\omega}_{\text{$s_1$ times}},
\underbrace{\bar \omega,\ldots,\bar \omega}_{\text{$s_2$ times}},
\underbrace{m,\ldots,m}_{\text{$r_1$ times}}, 
\underbrace{\bar m,\ldots,\bar m}_{\text{$r_2$ times}}),
\end{equation*}
possibly after changing
the order of the arguments of $T$. In principle, $\nu$ is a function
on $U\subset\St$. But under rotations of the frame, it gives rise to a
unique function on $\pi^{-1}(U)$ which changes along the fiber according to the transformation of the frame. This
function on $\Sth$ is denoted by the same symbol $\nu$ for simplicity.  In particular, its
dependence on the fiber coordinate $\rho$ is given by
\begin{align*}
  \nu&=e^{\i(r_1-r_2-s_1+s_2)\rho}
  T(\underbrace{\omega^*,\ldots,\omega^*}_{\text{$s_1$ times}},
  \underbrace{\bar \omega^*,\ldots,\bar \omega^*}_{\text{$s_2$ times}},
  \underbrace{m^*,\ldots,m^*}_{\text{$r_1$ times}},
  \underbrace{\bar m^*,\ldots,\bar m^*}_{\text{$r_2$ times}}
  )\\
  &=e^{\i(r_1-r_2-s_1+s_2)\rho}\nu^*,
\end{align*}
where we consider $\nu^*$ as independent of $\rho$ (because it is
defined with respect to the reference frame $(m^*,\bar m^*)$). We get
\begin{equation*}
  Y_3(\nu)=\i(r_1-r_2-s_1+s_2)\nu = \i s \nu,
\end{equation*}
where $s$ is the \highlight{spin-weight} introduced in
\cite{Penrose:1984tf}. Hence, \Eqref{eq:ethbracket} becomes
\[[\ethmath,\ethmath'](\nu)=-2 s \nu.\]

In summary, every quantity $\nu$ on $U\subset\St$ of spin-weight $s$ can be lifted to a smooth function $e^{\i s\rho}\cdot (\nu\circ\pi)$ on $\pi^{-1}(U)\subset \Sth$ (which we denote by the same symbol from now). Vice versa, every such function on $\pi^{-1}(U)$ pulls back to a function with spin-weight $s$ on $U$ along a smooth section over $U$. In the following, we will therefore often not distinguish between a function with spin-weight $s$ on \St and the corresponding function on \Sth.

In the case of the $2$-sphere with the standard round unit metric, we often consider the reference frame
\begin{equation}
  \label{eq:standardreferenceframe}
  m^*:=\frac 1{\sqrt 2}\left(\partial_{\vartheta}-\frac \i{\sin\theta}\partial_\varphi\right),
\end{equation}
and choose $U$ as the set of all points on $\St$ without the two poles $\vartheta=0,\pi$. As the smooth local section, we choose $\rho(\vartheta,\varphi)\equiv 0$.
Comparing this with the coordinate expressions above, we see that 
\begin{equation*}
   m^*=\pi^*(-\left.({Y_2} + \i{Y_1})\right|_{\rho=0})/\sqrt 2 =\pi^*(\left.\eth\right|_{\rho=0})/\sqrt 2.
\end{equation*}
Therefore, if $\nu$ is a function on $U$ with spin-weight $s$ and $\hat\nu$
(here, we exceptionally use two different symbols $\nu$ and
$\hat\nu$)
the corresponding function on $\pi^{-1}(U)$, then
\[\left.\eth\right|_{\rho=0}(\hat\nu)
=\left.(\partial_{\theta}-\frac \i{\sin\theta}\partial_\phi)\hat\nu+\i\cot\theta\partial_\rho\hat\nu\right|_{\rho=0}
=\left. (\sqrt 2\, m^*(\nu)-s\cot\vartheta\nu)\circ\pi\right|_{\rho=0}.\]
Under all these conditions, it makes sense therefore to simplify the notation and write
\begin{equation}
  \label{eq:ethm}
\eth(\nu)=\sqrt 2\, m^*(\nu)-s\nu\cot\vartheta,
\end{equation}
for a function on $\St$ with spin-weight $s$. In the same way, we obtain
\begin{equation}
\label{eq:ethm2}
\eth'(\nu)=\sqrt 2\,\mbar^*(\nu)+s\nu\cot\vartheta.
\end{equation}

\subsection{Spin-weighted spherical harmonics and decompositions via
Clebsch-Gordan Coefficients and Wigner \texorpdfstring{$3j$}{3j}-symbols.}
\label{sec:introCGw3j} 

In application of the Fourier-Galerkin (spectral) method to the solution of PDEs, products of spin-weighted spherical harmonics (SWSH) will be encountered. This motivates the exploration of a convenient method of treating these product terms --- which will result in the appearance of the Clebsch-Gordan (CG) series, calculation of which will be facilitated by the relation of coefficients in this series to the Wigner $3j$-symbols.

We proceed by first stating coordinate expressions for the well-known Wigner $\mathcal{D}$ matrices that form a basis for $L^2(SU(2))$, which will allow for the use of commonly encountered identities from the treatment of angular momentum in quantum mechanics \cite{Sakurai:1994modern}.

The Euler parametrization of a rotation can be written in terms of the Euler coordinates $\theta$, $\rho$, $\phi$ introduced above.  According to \cites{Goldberg:1967vm, Sakurai:1994modern} we have \begin{align}
  \mathcal{D}^l_{mn}(\rho,\theta,\phi) =& e^{\i m\rho} 
    \mathrm{d}^l_{mn}(\theta) e^{\i n\phi},\\\nonumber
  \mathrm{d}^l_{mn}(\theta)
  =& 
  \sum_{r=\max(0,m-n)}^{\min(l+m,l-n)}\left\{(-1)^{r-m+n} 
    \vphantom{\frac{\theta}{2}}\right.
  \frac{\sqrt{(l+m)!(l-m)!(l+n)!(l-n)!}}{r!(l+m-r)!(l-r-n)!(r-m+n)!}\\
    \label{eq:wigReddDef}
   &
   \left.\times\cos^{2l-2r+m-n}\frac{\theta}{2}\sin^{2r-m+n}\frac{\theta}{2} 
    \vphantom{\frac{\theta}{2}}\right\},
\end{align}
where $l\in\mathbb{N}_0:=\mathbb{N}\cup\{0\}$ (or $l\in\mathbb{N}_0+1/2$ for
spinorial quantities), $m,n\in\mathbb{Z}$, $|m|\leq l$, $|n|\leq l$. The
quantity $\mathrm{d}^l_{mn}$ is the reduced Wigner matrix element and satisfies
$\mathrm{d}^l_{mn}=\overline{\mathrm{d}^l_{mn}}$ together with the indicial
symmetry $\mathrm{d}^l_{mn}=(-1)^{m-n}\mathrm{d}^l_{nm}$. Upon introduction of
$(\vartheta,\varphi)$ on $\mathbb{S}^2$ as above we introduce\footnote{This choice is
standard and corresponds to a choice of smooth local section with
$\rho(\vartheta,\varphi)=0$, we assume this choice has been made henceforth,
unless otherwise specified -- see \S\ref{sec:geom}.}
the spin-weighted
spherical harmonics as
\begin{align}\label{eq:sphHarmDefnGold}
{}_{s}Y_{lm}(\vartheta,\varphi)
=&
\sqrt{\frac{2l+1}{4\pi}}\mathcal{D}_{sm}^l(0,\vartheta,\varphi),\\\label{eq:sphHarmDefnRedd}
=&
\sqrt{\frac{2l+1}{4\pi}}e^{\i m\varphi}\mathrm{d}^l_{sm}(\vartheta).
\end{align}
This fixes our convention to agree with~\cite{Penrose:1984tf}.
From Eq.~\eqref{eq:wigReddDef} and Eq.~\eqref{eq:sphHarmDefnRedd} we immediately observe a useful property the SWSH possess under complex conjugation:
\begin{equation}\label{eq:yslmconj}
\overline{{}_{s}Y_{lm}(\vartheta,\varphi)}=(-1)^{s-m}\,{}_{-s}Y_{l,-m}(\vartheta,\varphi).
\end{equation}
For later convenience, we compare (see also Eq.~(4.15.122)) of
\cite{Penrose:1984tf}) the algebraic action of the differential
operators $\eth,\,\eth'$ (cf. Eq.~\eqref{eq:ethm} and
Eq.~\eqref{eq:ethm2}) and their explicit coordinatizations
on~${}_sY_{lm}(\vartheta,\varphi)$:
\begin{align}\nonumber
\eth{}\,_sY_{lm}(\vartheta,\varphi) 
=&
(\sin\vartheta)^s\left(\partial_\vartheta-\i\csc\vartheta\partial_\varphi \right)\left[(\sin\vartheta)^{-s}\,{}_sY_{lm}(\vartheta,\varphi)\right]\\\label{eq:ethraco}
=&-\sqrt{(l-s)(l+s+1)} {}_{s+1}Y_{lm}(\vartheta,\varphi)\\\nonumber
\etp\,{}_sY_{lm}(\vartheta,\varphi) 
=&
(\sin\vartheta)^{-s}\left(\partial_\vartheta + \i\csc\vartheta\partial_\varphi \right)\left[(\sin\vartheta)^s\vartheta\,{}_sY_{lm}(\vartheta,\varphi)\right]\\\label{eq:ethloco}
=&
\sqrt{(l+s)(l-s+1)} {}_{s-1}Y_{lm}(\vartheta,\varphi)\\
\Delta_{\mathbb{S}^2}\,{}_sY_{lm}(\vartheta,\varphi)
=&
\frac{1}{2}\left(\eth\eth'+\eth'\eth \right)\left[{}_sY_{lm}(\vartheta,\varphi)\right]
=\left(s^2-l(l+1)\right)\,{}_sY_{lm}(\vartheta,\varphi)
\end{align}
For later reference, we also restate the orthonormality relation:
\begin{equation}\label{eq:sYlmOrth}
\left\langle {{}_sY_{l_1m_1}},\,{{}_sY_{l_2m_2}}\right\rangle
=
\int_0^{2\pi}\int_0^\pi
{}_sY_{l_1m_1}(\vartheta,\varphi) \overline{{}_sY_{l_2m_2}(\vartheta,\varphi)} 
\sin\vartheta\,d\vartheta\,d\varphi
=
\delta_{l_1l_2}\delta_{m_1m_2},
\end{equation}
which is directly inherited from the properties of the $\mathcal{D}$-matrices. Observe that in Eq.~\eqref{eq:sYlmOrth} orthonormality holds for functions of the same spin-weight.
We now describe a closed-sum decomposition for products such as:
\begin{align}\label{eq:spinWeightProd}
{}_{s_1}Y_{l_1,m_1}(\vartheta,\,\varphi)\,{}_{s_2}Y_{l_2,m_2}(\vartheta,\,\varphi)
=&
\sqrt{\frac{2l_{1}+1}{4\pi}}
\mathcal{D}_{s_{1}m_{1}}^{l_{1}}(0,\,\vartheta,\,\varphi)
\sqrt{\frac{2l_{2}+1}{4\pi}}
\mathcal{D}_{s_{2}m_{2}}^{l_{2}}(0,\,\vartheta,\,\varphi),
\end{align}
which together with the action of the $\eth,\,\eth'$ operators in Eq.\eqref{eq:ethraco} and Eq.\eqref{eq:ethloco} will form the basis of our spectral scheme.
The decomposition we seek is the so-called Clebsch-Gordan series which in bra-ket notation is given by \cite{Sakurai:1994modern}:
\begin{align}\nonumber
\mathcal{D}_{m_{1},n_{1}}^{l_{1}}\left(\rho,\,\theta,\,\phi\right)
\mathcal{D}_{m_{2},n_{2}}^{l_{2}}\left(\rho,\,\theta,\,\phi\right) 
= &
\sum_{l\in\Lambda}\left\{
\langle l_{1},\, l_{2};\, m_{1},\, m_{2}|\, l_{1},\, l_{2};\, l,\,(m_{1}+m_{2})\rangle\vphantom{\mathcal{D}^l_{(m)}}\right. \\\nonumber
  & 
\times\langle l_{1},\, l_{2};\, n_{1},\, n_{2}|\, l_{1},\, l_{2};\,l,\,(n_{1}+n_{2})\rangle \\\label{eq:ClebschGordanSeries}
&
\left.\mathcal{D}_{(m_{1}+m_{2}),(n_{1}+n_{2})}^{l}\left(\rho,\,\theta,\,\phi\right)\vphantom{\mathcal{D}^l_{(m)}}\right\},
\end{align}
where $\Lambda:=\{\max(|l_1-l_2|,\,|m_1+m_2|,\,|n_1+n_2|),\,\dots,\, l_1+l_2\}$. Note that each Clebsch-Gordan coefficient in the series is real, i.e. $\langle\cdot,\,\cdot\,;\,\cdot,\,\cdot|\,\cdot,\,\cdot\,;\,\cdot,\,\cdot \rangle\in\mathbb{R}$.
Define:
\begin{align}\nonumber
	\mathcal{A}_l(s_1,l_1,m_1;\,s_2,l_2,m_2)
	:=&
	\sqrt{\frac{(2l_1+1)(2l_2+1)}{4\pi}}\frac{1}{\sqrt{2l+1}}
	\left\langle l_1,s_1;l_2,s_2\right|\left. l_1,l_2;l,(s_1+s_2)\right\rangle\\\label{eq:CGauxA}
	 &
	\times \left\langle l_1,m_1;l_2,m_2 \right|\left.l_1,l_2;l,(m_1+m_2)\right\rangle.
\end{align}
Equation~\eqref{eq:CGauxA} together with Eq.~\eqref{eq:ClebschGordanSeries} thus provides us with the following decomposition of Eq.~\eqref{eq:spinWeightProd}:
\begin{equation}\label{eq:prodSoln}
	{}_{s_{1}}Y_{l_{1},m_{1}}(\vartheta,\,\varphi){}_{s_{2}}Y_{l_{2},m_{2}}(\vartheta,\,\varphi)
	=
	\sum_{l\in\Lambda'}\mathcal{A}_l(s_1,l_1,m_1;\,s_2,l_2,m_2)
	{}_{(s_{1}+s_{2})}Y_{l,(m_{1}+m_{2})}(\vartheta,\,\varphi),
\end{equation}
where $\Lambda':=\{\max(|l_1-l_2|,\,|s_1+s_2|,\,|m_1+m_2|),\,\dots,\,l_1+l_2\}$. Hence the product of two spin-weighted spherical harmonics may be decomposed into a finite linear combination of spin-weighted spherical harmonics with spin-weight equal to the sum of the original two spin-weights. Spectral decomposition of evolution equations will also require the following identity:
\begin{align}
\mathcal{I}\nonumber
=&
\int_0^{2\pi}\int_0^\pi
{}_{s_1}Y_{l_1m_1}(\vartheta,\varphi){}_{s_2}Y_{l_2m_2}(\vartheta,\varphi)\overline{{}_{(s_1+s_2)}Y_{l_3m_3}(\vartheta,\varphi)}
\sin\vartheta\,d\vartheta d\varphi,\\\label{eq:SpinnySpectroscope}
=&
\sum_{l\in\Lambda'}\mathcal{A}_l(s_1,l_1,m_1;\,s_2,l_2,m_2)\delta_{ll_3}\delta_{(m_1+m_2),m_3},
\end{align}
which may be obtained using Eq.~\eqref{eq:sYlmOrth} and Eq.~\eqref{eq:prodSoln}.

In the interest of efficient numerical calculations utilising the relation of the Clebsch-Gordan coefficients to the Wigner $3j$-symbols is prudent due to the convenient symmetry properties the latter possess \cite{Olver:2010nistHandbook}. We have
\begin{align}\label{eq:clebschGordWigRel}
\left(\begin{matrix}
j_{1} & j_{2} & j_{3}\\
m_{1} & m_{2} & m_{3}
\end{matrix}\right): &
=
\frac{(-1)^{j_1-j_2-m_3}}{\sqrt{2j_3+1}}\langle j_1,\, m_1;\, j_2,\, m_2|\, j_1,\, j_2;\, j_3,-m_3\rangle,
\end{align}
the non-negative quantities $\{j_1,\,j_2,\,j_3\}$ are known as \emph{angular momentum numbers} and may be integral or half-integral. The quantities $\{m_1,\,m_2,\,m_3\}$ are called the \emph{projective quantum numbers} and are given by $m_r=-j_r,\,-j_r+1,\,\dots,\,j_r-1,\,j_r$ where $r=1,2,3$.
Three further constraints are placed on the $j_i$ and $m_i$:
\begin{enumerate}[(WI)]
\item{$J:=j_1+j_2+j_3\in\mathbb{N}_0$;}
\item{$m_1+m_2+m_3=0$;}
\item{The \emph{triangle condition}: $|j_r-j_s|\leq j_t\leq j_r+j_s$ where $r,s,t$ is any permutation of $1,2,3$;}
\end{enumerate}
in the event these constraints fail to be satisfied the $3j$-symbol is set to $0$. The following symmetries will also be of use later
\begin{enumerate}[(SI)]
\item{Invariance under permutation of columns
\begin{align}\nonumber
 &\underbrace{
  \left(
   \begin{matrix}
    j_{1} & j_{2} & j_{3}\\
    m_{1} & m_{2} & m_{3}
   \end{matrix}
  \right) 
  =
  \left(
   \begin{matrix}
    j_{3} & j_{1} & j_{2}\\
    m_{3} & m_{1} & m_{2}
   \end{matrix}
  \right) 
  =
  \left(
   \begin{matrix}
    j_{2} & j_{3} & j_{1}\\
    m_{2} & m_{3} & m_{1}
   \end{matrix}
  \right)
  }_{\mbox{cyclic}}
  \\\nonumber
 =&\underbrace{
  (-1)^J
  \left(
   \begin{matrix}
    j_{3} & j_{2} & j_{1}\\
    m_{3} & m_{2} & m_{1}
   \end{matrix}
  \right)
  =
  (-1)^J
  \left(
   \begin{matrix}
    j_{1} & j_{3} & j_{2}\\
    m_{1} & m_{3} & m_{2}
   \end{matrix}
  \right)
  =
  (-1)^J
  \left(
   \begin{matrix}
    j_{2} & j_{1} & j_{3}\\
    m_{2} & m_{1} & m_{3}
   \end{matrix}
  \right)}_{\mbox{anticyclic}}
\end{align}

}
\item{Invariance under spatial inflection\footnote{Correction of Eq.~(2.8) of \cite{Rasch:2003efficientStorage}}
\begin{equation}
\nonumber
	\left(
			\begin{matrix}
				j_{1} & j_{2} & j_{3}\\
				m_{1} & m_{2} & m_{3}
			\end{matrix}
	\right)
	=
	(-1)^{J}\left(
		\begin{matrix}
			j_{1} & j_{2} & j_{3}\\
			-m_{1} & -m_{2} & -m_{3}
		\end{matrix}
	\right)
\end{equation}
}
\item{\emph{Regge symmetries}
\[
\begin{gathered}
	\left(
		\begin{matrix}
			j_{1} & j_{2} & j_{3}\\
			m_{1} & m_{2} & m_{3}
		\end{matrix}
	\right) 
	= 
	\left(
		\begin{matrix}
			j_{1} & \frac{1}{2}(j_{2}+j_{3}+m_{1}) & \frac{1}{2}(j_{2}+j_{3}-m_{1})\\
			j_{2}-j_{3} & \frac{1}{2}(j_{3}-j_{2}+m_{1})+m_{2} & \frac{1}{2}(j_{3}-j_{2}+m_{1})+m_{3}
		\end{matrix}
	\right)\\
	= 
	\left(
		\begin{matrix}
			\frac{1}{2}(j_{1}+j_{2}-m_{3}) & \frac{1}{2}(j_{2}+j_{3}-m_{1}) & \frac{1}{2}(j_{1}+j_{3}-m_{2})\\
			j_{3}-\frac{1}{2}(j_{1}+j_{2}+m_{3}) & j_{1}-\frac{1}{2}(j_{2}+j_{3}+m_{1}) & j_{2}-\frac{1}{2}(j_{1}+j_{3}+m_{2})
		\end{matrix}
	\right)
\end{gathered}
\]
}
\end{enumerate}
From the symmetries of the $3j$-symbols we recover the following symmetries on the $\mathcal{A}$ of Eq.~\eqref{eq:prodSoln} that we will make use of later:
\[
\begin{aligned}
\mathcal{A}_l(s_1,l_1,m_1;\,s_2,l_2,m_2)&=\mathcal{A}_l(s_2,l_2,m_2;\,s_1,l_1,m_1),\\
\mathcal{A}_l(s_1,l_1,m_1;\,s_2,l_2,m_2)&=(-1)^{l+l_1+l_2-2(s_1+s_2)}\mathcal{A}_l(-s_1,l_1,m_1;\,-s_2,l_2,m_2),\\
\mathcal{A}_l(s_1,l_1,m_1;\,s_2,l_2,m_2)&=(-1)^{l+l_1+l_2}\mathcal{A}_l(s_1,l_1,-m_1;\,s_2,l_2,-m_2).
\end{aligned}
\]
When performing a decomposition such as in Eq.~\eqref{eq:prodSoln} numerically, it can be convenient to embed the above symmetries directly into the summation process. Furthermore, there also exist efficient storage schemes for $3j$-symbols, utilizing symmetries such as (SI-SIII) \cite{Rasch:2003efficientStorage} -- this allows for precalculation of all required $3j$-symbols and evaluating Eq.~\eqref{eq:prodSoln} in this manner may be more efficient under certain circumstances.


\section{Numerical method}

\subsection{Spectral transformation}
\label{sec:specXform}

In this section we briefly describe the numerical implementation of Huffenberger and Wandelt's spectral algorithm \cite{Huffenberger:2010hh} that will allow for the decomposition of an integer spin-weighted function ${}_sf\in L^2(SU(2))$ in terms of the SWSH of Eq.~\eqref{eq:sphHarmDefnGold}.  Numerical calculations must be limited to a finite grid, hence the decomposition must be truncated at a maximal harmonic (band-limit) $L$.  In terms of this band limit the algorithm has $\mathcal{O}(L^3)$ complexity which is achieved by exploiting a smooth periodic extension of the data to the $2$-torus so that existing Fast Fourier Transform (FFT) methods can be used.

Consider a function ${}_{s}\hat{f}\in L^2(SU(2))$. By the Peter-Weyl theorem for compact groups \cite{Sugiura:1990vj} we have
\begin{equation}\nonumber
\lim_{L\rightarrow\infty}\left\Vert
{}_{s}\hat{f}(\theta,\phi,\rho)
-\sum^L_{l=|s|}\sum^{l}_{m=-l}
{}_sa_{lm} \sqrt{\frac{2l+1}{4\pi}} \mathcal{D}^l_{sm}(\rho,\theta,\phi)
\right\Vert_2=0.
\end{equation}
Fixing the fibre coordinate as $\rho=0$ and performing the map Eq.~\eqref{eq:localHopf} we then define ${}_sf(\vartheta,\varphi):=\left.{}_s\hat{f}(\theta,\phi,\rho)\right|_{\rho=0}$ and arrive at the standard expansion
\begin{equation}\label{eq:funcs2sphexpInf}
{}_{s} f(\vartheta,\varphi)
=
\lim_{L\rightarrow \infty} \sum^L_{l=|s|}\sum^{l}_{m=-l} {{}_sa_{lm}} \,{{}_sY_{lm}(\vartheta,\varphi)}
\end{equation}
Henceforth, we work with the band-limited expression:
\begin{equation}\label{eq:funcs2sphTrunc}
{}_{s} f(\vartheta,\varphi)
=
\sum^L_{l=|s|}\sum^{l}_{m=-l} {{}_sa_{lm}} \,{{}_sY_{lm}(\vartheta,\varphi)},
\end{equation}
where it is assumed that the function being decomposed may be completely expressed by a finite linear combination of the basis functions.

\subsubsection{Forward transformation}
\label{sec:algofwdXform}

We now describe the algorithm for evaluation of the forward transform $\mathcal{F}: {}_sf\mapsto ({}_sa_{lm})$. As a first step introduce the notation $\Delta_{m n}^l:=\mathrm{d}_{m n}^l\left(\pi/2\right)$ which allows for the rewriting of Eq.~\eqref{eq:wigReddDef} as~\cite{Risbo:1996iy} 
\begin{equation}
\label{eq:wignerdDecDelta} 
\mathrm{d}_{m n}^l(\vartheta)=\i^{m-n}\sum_{q=-l}^{l}\Delta_{q m}^le^{-iq\vartheta}\Delta_{q n}^l, \end{equation} 
following from a factoring of rotations \cite{Trapani:2006he}. The details of how the $\Delta$ elements are calculated together with their symmetry properties are given in \S\ref{sec:algodelCalc}.

Define the functional:
\begin{equation}\label{eq:ImnQuadPr}
I_{mn}\left[{}_sf(\vartheta,\varphi) \right]:=\int^{2\pi}_0\int^\pi_0 e^{-\i m\vartheta}e^{-\i n\varphi}{}_s f(\vartheta,\varphi) \sin\vartheta \,d\vartheta d\varphi.
\end{equation}
Equation~\eqref{eq:sphHarmDefnRedd} together with Eqs.~(\ref{eq:funcs2sphexpInf}-\ref{eq:ImnQuadPr}) lead to
\[
\begin{aligned}\label{eq:sphcoeffIntegral2}
	{}_sa_{lm}
	&=\i^{s-m}\sqrt{\frac{2l+1}{4\pi}}\sum^{l}_{q=-l} \Delta^l_{qm} I_{qm} \Delta^l_{qs} \end{aligned} \] We now wish to evaluate the expression for $I_{mn}$. This may be done exactly, by extension of the function ${}_sf$ to the $2$-torus $\mathbb{T}^2$, which will permit the application of the standard $2$D Fourier-transform. Although this requires computation of points outside the domain of interest, the corresponding increase in the speed of performing the calculation (for large $L$) and favourable (spectral)-convergence offered by this method compensates for the increased computational effort.  Define the extended function 
\begin{equation}
\nonumber
	{}_sF\left(\vartheta,\varphi\right):=
	\begin{cases}
		{}_sf(\vartheta,\varphi) & \vartheta\leq\pi\\
		{}_sf(\pi-\vartheta,\varphi) & \vartheta>\pi
	\end{cases}
\end{equation}
where $\vartheta$ now takes values in $[0,2\pi)$. Clearly this does not change the value of $I_{mn}$ because extension of the function leaves its value unchanged within the domain of integration as defined in Eq.~\eqref{eq:ImnQuadPr}. The periodic extension is chosen by forming a linear combination of ${}_s Y_{lm}(\vartheta,\varphi)$ and examining symmetries using the defining relations of Eq.~\eqref{eq:sphHarmDefnRedd} and Eq.~\eqref{eq:wignerdDecDelta}. As the periodically extended function ${}_s F(\vartheta,\varphi)$ now possesses $2\pi$ periodicity in both arguments, it may be written as the two dimensional, band-limited Fourier sum
\begin{equation}\nonumber
{}_s F(\vartheta,\varphi)=\sum_{k,n=-L}^L F_{kn}\exp\left(\i k\vartheta\right)\exp\left(\i n\varphi\right).
\end{equation}
Substitution of this equation into Eq. \eqref{eq:ImnQuadPr} yields:
\begin{subequations}
 \begin{align}
  I_{m'm} & = 
  						\sum_{k,n=-L}^{L} F_{kn}\left[\int_{0}^{2\pi}\exp\left(\i(n-m)\varphi\right)d\varphi\right]
  						\left[\int_{0}^{\pi}\exp\left(\i(k-m')\vartheta\right)\sin\vartheta\,d\vartheta\right],\\ 
					& = 2\pi\sum_{k=-L}^{L}F_{km} w(k-m'), \label{eq:weightConvo}
 \end{align}
\end{subequations}
where $w(p)$ for $p\in\mathbb{Z}$ is given by
\footnote{(B5) of \cite{Huffenberger:2010hh} contains an error.}
\begin{equation}\nonumber
\label{eq:quadWeightDefn}
	w(p)=\int_{0}^{\pi}\exp\left(\i p\vartheta\right)\sin\vartheta\,d\vartheta=
	\begin{cases}
		2/(1-p^{2}) & p\;\mathrm{even}\\
		0 & p\;\mathrm{odd},\, p\neq\pm1\\
		\pm \i\pi/2 & p=\pm1.
              \end{cases} \end{equation} 
Equation~\eqref{eq:weightConvo} shows that $I_{m'm}$ is proportional to a discrete convolution in spectral space. By the convolution theorem, this implies that we may consider instead the inverse transform of $w(p)$ which maps the function back to its spatial representation $w_r$. Performing point-wise multiplication with $2\pi\,{}_sF$ and transforming the result will yield $I_{m'm}$. If the desired number of samples of the function $_sf(\vartheta,\varphi)$ over $\vartheta$ and $\varphi$ on $\mathbb{S}^2$ is to be $N_\vartheta$ and $N_\varphi$ respectively, then for the number of samples for the extended function $_sF(\vartheta,\varphi)$ we take to be $N_\vartheta'=2(N_\vartheta-1)$ and $N_\varphi$. The spatial sampling intervals are given by $\Delta\vartheta=\frac{2\pi}{N_\vartheta'}$ and $\Delta\varphi=\frac{2\pi}{N_\varphi}$. Note that in order to satisfy the Nyquist condition, we must take $N_\vartheta=2(L+2)$ and $N_\varphi=2(L+2)$, where $L$ is the harmonic that the function ${}_s f(\vartheta,\varphi)$ is band-limited to. With the stated sampling, the quadrature weights may be written as 
\begin{equation} 
\label{eq:quadWeightxForm} w_r(q'\Delta\vartheta)=\sum^{N_\vartheta'/2-1}_{p=-N_\vartheta'/2}\exp\left(-\i pq'\Delta\vartheta \right)w\left(p\right).  
\end{equation} 
Upon performing a linear phase shift in Eq. \eqref{eq:quadWeightxForm} we finally arrive at the expression 
\begin{equation}
\nonumber
	I_{m'm}=
	\frac{2\pi}{N'_\vartheta N_\varphi}
	\sum^{N'_\vartheta-1}_{q'=0}
	\sum^{N_\varphi-1}_{q=0}
	\exp\left(-\i m'q'\Delta\vartheta \right)
	\exp\left(-\i mq\Delta\varphi\right)
	w_r\left(q'\Delta\vartheta\right)
	{}_{s}F(q'\Delta\vartheta,q\Delta\varphi),
\end{equation}
which may be evaluated using a $2$-dimensional FFT. We note that if many
transformations with the same band-limits are to be performed, the weights can
be pre-calculated.

Overall, the complexity of the outlined algorithm is $\mathcal{O}(L^3)$. Two
further linear improvements in execution speed are possible. The first reduces
the total computation time of ${}_s a_{lm}$ by a factor of~$2$.
Equation~\eqref{eq:sphcoeffIntegral2} together with the symmetries of
$\Delta^l_{mn}$ described in \S\ref{sec:algodelCalc} allows for:
\begin{equation}\nonumber
\sum_{q=-l}^l\Delta^l_{qm}I_{qm}\Delta^l_{q(-s)}=\sum^{l}_{q=0}\Delta^l_{qm} K_{qm} \Delta^l_{q(-s)},
\end{equation}
where
\begin{equation}\nonumber
K_{qm}:=
	\begin{cases}
		I_{0m} & \mbox{if }q=0\\
		I_{qm}+(-1)^{m+s}I_{(-q)m} & \mbox{if }q>0.
	\end{cases}
\end{equation}
A second improvement (also by a factor of~$2$) is possible if the function ${}_{s}f$ being analysed is real. Here the speedup is due to the FFT, where real input results in Hermitian output.

\subsubsection{Backward tranformation}
\label{sec:algobwdXform}

We now describe the algorithm for evaluation of the backward (inverse) transform $\mathcal{F}{}^{-1}:({}_sa_{lm})\mapsto{}_sf$.  The backward spherical harmonic transform maps the expansion coefficients ${}_sa_{lm}$, for $|s|\leq l \leq L$, to a function on $\mathbb{S}^2$.  Because we are working with band-limited functions we can, at least in theory, perfectly reconstruct the original function. To this end, Eq.~\eqref{eq:funcs2sphTrunc} must be evaluated. As the inverse transform does not contain integrals, issues of quadrature accuracy do not arise.

Define the functional:
\begin{equation}\label{eq:JmnC}
J_{mn}\left[{}_s a_{ln} \right]:= \i^{s-n} \sum_{l=|s|}^L \sqrt{\frac{2l+1}{4\pi}} \Delta^l_{(-m)s}\, {}_sa_{ln}\, \Delta^l_{(-m)n}.
\end{equation}
Substitution of Eq.~\eqref{eq:sphHarmDefnRedd} together with Eq.~\eqref{eq:wignerdDecDelta} and Eq.~\eqref{eq:JmnC} in Eq.~\eqref{eq:funcs2sphTrunc} leads to
\begin{equation}
\label{eq:bwdFSH}
	{}_sf(\vartheta,\varphi) 
	=
	\sum_{m=-N_\vartheta'}^{N_\vartheta'}\sum_{n=-N_\varphi}^{N_\varphi} e^{\i m\vartheta}e^{\i n\varphi} J_{mn}.
\end{equation}
Evaluation of~\eqref{eq:bwdFSH} results in ${}_sf(\vartheta,\varphi)$ sampled on $\mathbb{T}^2$. As we require the function on $\mathbb{S}^2$ we may truncate the output at $\vartheta=\pi$ discarding all data for $\vartheta>\pi$. Evaluating~\eqref{eq:bwdFSH} scales as $\mathcal{O}(L^3)$, just as when performing the forward transformation.

Taking into account the symmetries of the $\Delta^l_{mn}$ matrices provided by Eqs.~ (\ref{eq:DelIndSym}) leads to an analogous halving of the number of operations required for the evaluation of Eq.~\eqref{eq:bwdFSH}, as in the case of the forward transform.
Similarly, a further speedup is possible if the input data to the FFT library is Hermitian.

\subsubsection{Calculation of \texorpdfstring{$\Delta$}{Delta} elements}\label{sec:algodelCalc}
In this section we follow \cite{Trapani:2006he} and briefly outline an
efficient recursive method for computing the $\Delta^l_{mn}$ that appear upon
decomposition of the Wigner $d$-matrices (Eq.~\eqref{eq:wignerdDecDelta}) when
constructing the transformations in \S\ref{sec:algofwdXform} and
\S\ref{sec:algobwdXform}. 
It can be seen directly from Eq.~\eqref{eq:wigReddDef} that the elements $\Delta_{m n}^l$ have the following symmetries
\begin{equation}
\label{eq:DelIndSym}
\begin{aligned}
\Delta^l_{(-m)n}=(-1)^{l+n}\Delta^l_{mn},\\
\Delta^l_{m(-n)}=(-1)^{l-m}\Delta^l_{mn},\\
\Delta^l_{mn}=(-1)^{n-m}\Delta^l_{nm},
\end{aligned}
\end{equation}
where the $\Delta^l_{m n}$ are combinatorial expressions purely dependent on the choice of indices $l,\,m$ and $n$.

Suppose we require all possible $\Delta^l_{mn}$ up to a maximum $L$. Due to the symmetries of Eqs.~(\ref{eq:DelIndSym}), only a subset of all allowable indices need be calculated. For each choice of $l$ we restrict the indices for which $\Delta^l_{mn}$ is calculated to the set $\{(m,\,n)\,|\,m\in\{0,\,1,\,\dots l\}\,\mbox{and}\, n\in\{0,\,1,\,\dots l\}\}$.
\par We implement the \emph{Trapani-Navaza} (TN) algorithm as follows:
\begin{enumerate}[(TN I)]
\item{Initialise $\Delta^0_{00}=1$;}
\item{Iterate with
\begin{equation}\nonumber
  \Delta^l_{0l}=-\sqrt{\frac{2l-1}{2l}}\Delta^{(l-1)}_{0(l-1)};
\end{equation}
}
\item{Iterate with
\begin{equation}\nonumber
  \Delta^l_{ml}=\sqrt{\frac{l(2l-1)}{2(l+m)(l+m-1)}}\Delta^{(l-1)}_{(m-1)(l-1)};
\end{equation}
}
\item{
  Iterate with
  \begin{equation}\label{eq:DelTN3tunst}
    \Delta^l_{mn}=
      \begin{cases}
        \frac{2m}{\sqrt{(l-n)(l+n+1)}}\Delta^l_{m(n+1)}=\sqrt{\frac{2}{l}}m\Delta^l_{m(n+1)} & n=l-1,\\
        \frac{2m}{\sqrt{(l-n)(l+n+1)}}\Delta^l_{m(n+1)}-\sqrt{\frac{(l-n-1)(l+n+2)}{(l-n)(l+n+1)}}\Delta^l_{m(n+2)} & n<l-1,
      \end{cases}
      ;
  \end{equation}
}
\item{
Use symmetries (i.e. Eqs. (\ref{eq:DelIndSym})) to find the remaining $\Delta^l_{mn}$.
}
\end{enumerate}
Note that the $\Delta^l_{mn}$ may be viewed as having a square pyramidal lattice structure with TN I-IV being a calculation of an octant subset -- TN V then allows for all values to be found: TN I corresponds to the apex $(0,0,0)$; TN II corresponds to a tangent of values, a descent from the apex through the points $(l,0,l)$; TN III corresponds to calculation of a right-angle, triangular lattice of surface values $(l,m,l)$; TN IV corresponds to calculation of interior points $(l,m,n)$ constrained to lie in the set $\{(m,\,n)\,|\,m\in\{0,\,1,\,\dots l\}\,\mbox{and}\, n\in\{0,\,1,\,\dots l\}\}$; TN V allows for recovery of all valid values of $\Delta^l_{mn}$ (those outside the octant subset but within the square pyramid) up to the chosen limit $L$.

One advantage of the TN algorithm, is that it is well suited to
parallelization. Performing steps TN I-III and retaining the results then
allows for the trivially parallelizable TN IV to be performed on multiple
threads, as required. A disadvantage, however, is instability for large values
of $L$ (for our implementation $\geq 2595$). This issue, together with our
proposal to correct for it
without loss of efficiency will be described in~\S\ref{sec:wdelstab}.

\subsection{Computation of Clebsch-Gordan Coefficients}
\label{sec:CGw3jNumProc}

In the solution of PDEs, product terms of SWSHs arise which may be decomposed
as in Eq.~\eqref{eq:prodSoln}. Two obvious paths are open to us:
the first option is to implement a pseudo-spectral approach; extracting the
requisite $\mathcal{A}_l$ by transforming two sets of appropriately seeded ${}_{s}a_{lm}$
coefficient sets to functions, pointwise multiplying, then transforming back.
We shall revisit this approach in \S\ref{sec:numMax}. 

The second option, which we take, is direct computation of $A_l$ factors by calculation of Clebsch-Gordan coefficients (or Wigner-$3j$-symbols by Eq.~\eqref{eq:clebschGordWigRel}). To this end we now describe an exact three-term linear recursive algorithm for calculating the Wigner-$3j$-symbols due to \cite{Schulten:1975:exactW3j}.  The scheme we describe is numerically stable for `small' ($<100$) values of $j$ and $m$. However, the possible occurrence of numerical loss of significance as well as overflow leads us to subsequently refine our approach.  We apply the general conversion of a three term linear recursion relation into a hybrid recursion relation, given by Luscombe and Luban \cite{Luscombe:1998:SimplifiedW3jNL}, to the recursion given in \cite{Schulten:1975:exactW3j}.  Our implementation extends the mentioned schemes to cover calculation of $3j$-symbols of \emph{both} integer and half-integer angular momentum numbers and projective quantum numbers.

Our aim is the simultaneous generation of
\begin{equation}\label{eq:connectw3j}
	w(j_1):=\left(
	\begin{matrix}
		j_{1} & j_{2} & j_{3}\\
		m_{1} & m_{2} & m_{3}
	\end{matrix}
	\right)\qquad \text{with fixed } j_2,\,j_3,\,m_1,\,m_2,\,m_3
\end{equation}
for all $j_1\in\left\{j_{1_\mathrm{min}},\,\dots,\,j_{1_\mathrm{max}}\right\}$ where $j_{1_\mathrm{min}}=\mathrm{max}\{|j_2-j_3|,\,|m_1|\}$, $j_{1_{\mathrm{max}}}=j_2+j_3$, and where, in addition, $j_1$ is subject to the constraints discussed in \S\ref{sec:introCGw3j}. Consider Eq.~\eqref{eq:connectw3j}; if $j_1-1,j_1,j_1+1$ each provide a valid $3j$ symbol\footnote{Terms involving $j_1$ that are outside the range of validity are set to $0$ in the recursion relation.} then the symbols may be connected via the following three-term linear recursion relation \cite{Schulten:1975:exactW3j}:
\begin{equation}\label{eq:w3jrecursionReln}
j_{1}A(j_{1}+1)w(j_1+1)+B(j_{1})w(j_1)+(j_{1}+1)A(j_{1})w(j_1-1)=0,
\end{equation}
where
\begin{align}\label{eq:w3jAcoeff}
	A(j_{1}) 
	&:= 
	\left[j_{1}{}^{2}-(j_{2}-j_{3})^{2}\right]^{1/2}
	\left[(j_{2}+j_{3}+1)^{2}-j_{1}{}^{2}\right]^{1/2}
	\left[j_{1}{}^{2}-m_{1}{}^{2}\right]^{1/2},\\\label{eq:w3jBcoeff}
	B(j_{1})
	&:=
	-(2j_{1}+1)
	\left[j_{2}(j_{2}+1)m_{1}-j_{3}(j_{3}+1)m_{1}-j_{1}(j_{1}+1)(m_{3}-m_{2})\right].
\end{align}
The normalisation condition
\begin{equation}\label{eq:normw3j}
	\sum_{j_{1}=j_{1_{\mathrm{min}}}}^{j_{1_{\mathrm{max}}}}(2j_{1}+1)
	\left(
		\begin{matrix}
		j_{1} & j_{2} & j_{3}\\
		m_{1} & m_{2} & m_{3}
		\end{matrix}
	\right)^{2}=1,
\end{equation}
together with the phase convention
\begin{equation}\label{eq:phasew3j}
	\mathrm{sign}
	\left\{
		\left(
			\begin{matrix}
				j_{1_\mathrm{max}} & j_{2} & j_{3}\\
				m_{1} & m_{2} & m_{3}
			\end{matrix}
		\right)
	\right\} =(-1)^{j_{2}-j_{3}-m_{1}}
\end{equation}
allows for the determination of the family of $3j$-symbols of Eq.~\eqref{eq:connectw3j}.

\begin{figure}[ht]
	\centering
	\includegraphics{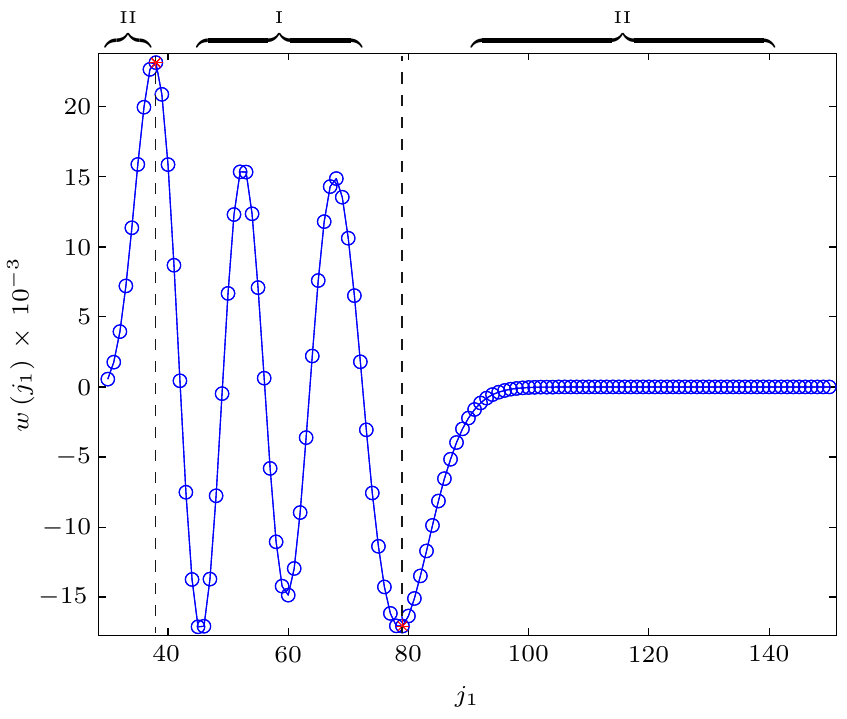}
	\caption{
	(Colour online) Typical functional form for a family of $3j$-symbols as specified by Eq.~\eqref{eq:connectw3j}. 
	Parameters: $j_2=90,$ $j_3=60;$ $m_1=-15,$ $m_2=70,$ $m_3=-55;$.
	Classical region denoted by $\mathrm{I}$, Non-classical regions denoted by $\mathrm{II}$.
	(Red stars) Left boundary of classical region $j_{1_-}=38$; Right boundary of classical region $j_{1_+}=79$.
	There are 26 orders of magnitude difference between the largest and smallest values of $w(j_1)$ in this family.
	Computed using hybrid algorithm.
	}
	\label{fig:wig3jRecur}
\end{figure}

The range of $j_1$ in Eq.~\eqref{eq:connectw3j} is divided into a `classical'
and two complementary `non-classical' regions. A classical region is defined as
the set of $\{j_1,j_2,j_3\}$ and $\{m_1,m_2,m_3\}$ for which it is possible to
construct a vector diagram which physically corresponds to the coupling
of angular momentum -- for more details, see
\cite{Luscombe:1998:SimplifiedW3jNL}. For our purposes it is sufficient to
consider the non-classical region as values of $|w(j_1)|$ that monotonically
decrease as $j_1\rightarrow j_{1_\mathrm{min}}$ and $j_1\rightarrow
j_{1_\mathrm{max}}$.
Consider Fig.~\ref{fig:wig3jRecur} where we illustrate the typical functional form
of $w(j_1)$. In the classical region (denoted $\mathrm{I}$) the amplitude of
$w(j_1)$ oscillates about $0$. Within this
region $w(j_1)$ may evaluate to
$0$ for specific choices of $j_1$. The left boundary of the classical
region is denoted by $j_{1_-}$, the right boundary is denoted by $j_{1_+}$. In
the non-classical regions (denoted $\mathrm{II}$) $|w(j_1)|$ monotonically
decays to zero as the boundaries $j_{1_\mathrm{min}}$ and $j_{1_\mathrm{max}}$
are approached.

In order to achieve numerical stability, the recursion relation of
Eq.~\eqref{eq:w3jrecursionReln} that is used to generate the quantities
$w(j_1)$ must be performed in the direction of increasing $|w(j_1)|$. More
explicitly, if the desired solution of a recurrence relation such as
Eq.~\eqref{eq:w3jrecursionReln}, is monotonically decreasing (as is the case if
one iterates from classical to non-classical region) then the other, linearly
independent solution is monotonically increasing. Thus numerical
round-off in the calculation of a decaying solution of the recurrence
relation triggers the growth of the unwanted, linearly independent,
diverging solution \cite{Luscombe:1998:SimplifiedW3jNL}. This discussion
implies that instead, we should initialise Eq.~\eqref{eq:w3jrecursionReln} at
the boundaries $j_{1_\mathrm{min}}:=\min(j_1)$ (left-to-right recursion) and
$j_{1_\mathrm{max}}:=\max(j_1)$ (right-to-left recursion) and proceed toward
the classical region $\mathrm{I}$.

From Eq.~\eqref{eq:w3jAcoeff} observe that $A(j_{1_\mathrm{min}})=0$ and $A(j_{1_\mathrm{max}}+1)=0$. The recursion relations at the boundaries, are given by
\[
\begin{aligned}
  B(j_{1_{\mathrm{min}}})w(j_{1_{\mathrm{min}}})+j_{1_{\mathrm{min}}}A(j_{1_{\mathrm{min}}}+1)w(j_{1_{\mathrm{min}}}+1) & =  0;\\
  B(j_{1_{\mathrm{max}}})w(j_{1_{\mathrm{max}}})+(j_{1_{\mathrm{max}}}+1)A(j_{1_{\mathrm{max}}})w(j_{1_{\mathrm{max}}}-1)
  & =  0.
\end{aligned}
\]
Note that the specification of \emph{one} initial parameter at each
boundary is sufficient to start the recursion (Eq.~\eqref{eq:w3jrecursionReln})
in each direction.

As the initial choice of parameter for both the left and right recursions is

arbitrary, the set of values generated by each recursion will be in error by
factors $c_L$ and $c_R$ respectively; explicitly
\[
\begin{aligned}
  c_{L}w(j_{1_{\mathrm{min}}});\; c_{L}w(j_{1_{\mathrm{min}}}+1);\;\dots;\; c_{L}w(j_{1_{mid}}) & \qquad  \mbox{(left-to-right recursion)};\\
  c_{R}w(j_{1_{\mathrm{max}}});\; c_{R}w(j_{1_{\mathrm{max}}}-1);\dots;\; c_{R}w(j_{1_{\mathrm{mid}}}) & \qquad  \mbox{(right-to-left recursion),}
\end{aligned}
\]
where the recursion is terminated at a common, final $j_1$ value of $j_{1_\mathrm{mid}}$. The left and right recursions must match \footnote{In order to achieve a robust implementation, one performs left and right recursions until several common $j_1$ values are achieved. This allows for the avoidance of singular terms in the classical region.} at $j_{1_\mathrm{mid}}$, which implies the constraint $c_L w(j_{1_\mathrm{mid}})=c_R w(j_{1_\mathrm{mid}})$. We rescale the left recursion by $c_R/c_L$ 
and determine $c_R$ from the normalisation condition~\eqref{eq:normw3j}. Finally, upon application of the phase condition~\eqref{eq:phasew3j}, we have generated all valid $w(j_1)$ as specified by~\eqref{eq:connectw3j}.

The algorithm just discussed suffers both from numerical loss of significance as well as overflow. This is due to the large variation in $|w(j_1)|$ for particular parameter choices; for example in the family of symbols shown in Fig.~\ref{fig:wig3jRecur} there are 26 orders of magnitude difference between the largest and smallest values of $|w(j_1)|$. The loss of significance/overflow is often mitigated by rescaling of iterates in the recursion. Alternatively, the use of a two-term nonlinear recursion may be employed \cite{Luscombe:1998:SimplifiedW3jNL}. To this end, we instead work with ratios of successive $w(j_1)$ terms.

Define the nonlinear left-to-right recursion
\begin{equation}\label{eq:nonlinleftrec}
s(j_1):=\frac{w(j_1)}{w(j_1+1)} = \frac{-j_1A(j_1+1)}{B(j_1)+(j_1+1)A(j_1)s(j_1-1)},\quad j_1\geq j_{1_\mathrm{min}}+1,
\end{equation}
and the nonlinear right-to-left recursion
\begin{equation}\label{eq:nonlinrightrec}
r(j_1):=\frac{w(j_1)}{w(j_1-1)} = \frac{-(j_1+1)A(j_1)}{B(j_1)+j_1A(j_1+1)r(j_1+1)},\quad j_1\leq j_{1_\mathrm{max}}-1;
\end{equation}
Since $A(j_{1_\mathrm{max}})=0$ and $A(j_{1_\mathrm{min}}+1)=0$ the initial values
\[
\begin{aligned}
s(j_{1_\mathrm{min}}+1) & =  \frac{-(j_{1_\mathrm{min}}+1)A(j_{1_\mathrm{min}}+2)}{B(j_{1_\mathrm{min}}+1)};\\
r(j_{1_\mathrm{max}}-1) & =  \frac{-j_{1_\mathrm{max}}A(j_{1_\mathrm{max}}-1)}{B(j_{1_\mathrm{max}}-1)},
\end{aligned}
\]
are known. The numerical advantage of making the transformations of Eq.~\eqref{eq:nonlinleftrec} and Eq.~\eqref{eq:nonlinrightrec}, is that $s(j_1)$ and $r(j_1)$ maintain values of order unity, throughout the recursion (i.e., at each iterate) thus avoiding significance/overflow issues. A disadvantage introduced by the nonlinear scheme is that Eq.~\eqref{eq:nonlinleftrec} and Eq.~\eqref{eq:nonlinrightrec} are poorly defined for values of $j_1$ where $w(j_1)=0$. However, as this occurs only in the classical region $\mathrm{I}$, this motivates the consideration of a hybrid scheme -- a combination of nonlinear (in the nonclassical regions $\mathrm{II}$) and linear (in the classical region $\mathrm{I}$). 

In order to split recursion schemes between the two methods, the location of the left $j_{1_-}$ and right $j_{1_+}$ boundaries must be known.  The precise choice of $j_{1_-}$ and $j_{1_+}$ is not crucial; the essential point is that nonlinear iteration is terminated near the boundary, such that no zero values of $w(j_1)$ in the classical region are encountered.  As $|w(j_1)|$ monotonically decreases in the nonclassical regions and is known to be non-zero there, a simple algorithmic method of determining the aforementioned values is to begin iteration with Eq.~\eqref{eq:nonlinleftrec} and Eq.~\eqref{eq:nonlinrightrec}, comparing the magnitude of two consecutive iterates until a local maximum in $|w(j_1)|$ is achieved.

The coefficients as determined by the nonlinear scheme in the nonclassical region are hence computed using
\[
\begin{aligned}
	w(j_{1_{-}}-k) & = w(j_{1_{-}})\prod_{p=1}^{k}s(j_{1_{-}}-p),\quad1\leq k\leq j_{1_{-}}-j_{1_{\mathrm{min}}};\\
	w(j_{1_{+}}+k) & = w(j_{1_{+}})\prod_{p=1}^{k}r(j_{1_{+}}+p),\quad1\leq k\leq j_{1_{\mathrm{max}}}-j_{1_{+}}.
\end{aligned}
\]
Once the coefficients in the nonclassical regions are known, we may perform iteration using the three-term linear recurrence relation in the classical region using $\{w(j_{1_{-}}-1),\,w(j_{1_{-}})\}$ and $\{w(j_{1_{+}}),\,w(j_{1_{+}}+1)\}$ as initial values in Eq.~\eqref{eq:w3jrecursionReln} for the left and right recursion respectively. The left and right hybrid recursion schemes must match at a common $j_{1_\mathrm{mid}}$, upon appropriate rescaling, and normalisation (Eq.~\eqref{eq:normw3j}) we finally apply the phase convention of Eq. \eqref{eq:phasew3j}. This procedure allows for the determination of an entire family of $w(j_1)$ as stated in Eq.~\eqref{eq:connectw3j}.

If the coefficients are required at several stages of a calculation, it is
generally more efficient to perform a recursive pre-computation, exploiting the
symmetries (SI-SIII) (see \S\ref{sec:introCGw3j}) in order to minimise memory
usage. We implement this approach in our code where feasible, for details see
\cite{Rasch:2003efficientStorage}.

\begin{figure}[ht]
\centering
\begin{subfigure}{\includegraphics[width=.44\linewidth] {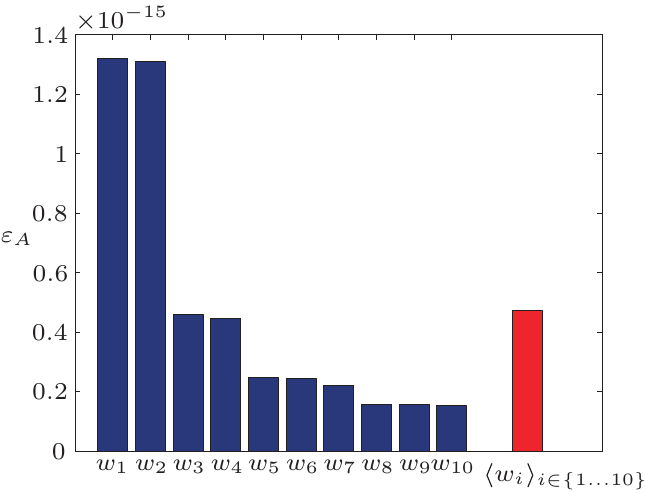}
   \label{fig:wignerError_sub1}
 }%
\end{subfigure}\hfill
 \begin{subfigure}{\includegraphics[width=.54\linewidth]{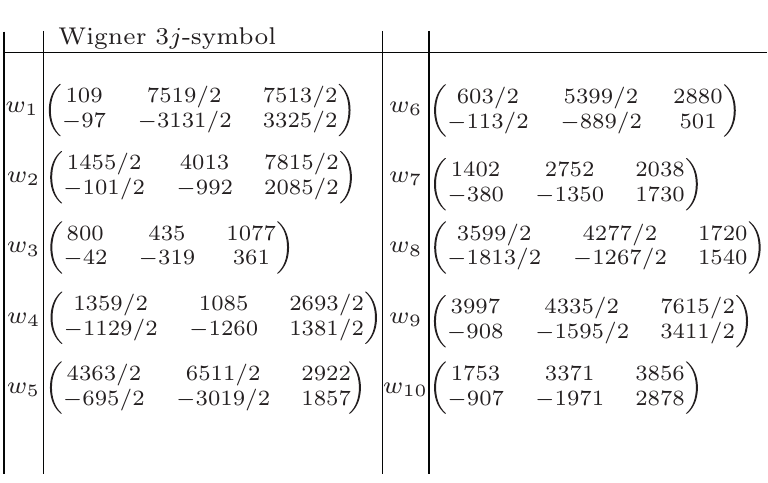}
   \label{fig:wignerError_sub2}
 }%
\end{subfigure}%
\caption{\label{fig:wig3jerrorFig}
	(Colour online) An error comparison between the hybrid-recursive algorithm (see \S\ref{sec:CGw3jNumProc}) and a direct symbolic calculation. 
In this calculation, random sets of valid Wigner $3j$-symbols were generated with possible parameters between $j_\mathrm{min}=0$ and $j_\mathrm{max}=4048$. Upon generation of an invalid random combination of $\{j_i\}$ and $\{m_i\}$, the parameters were discarded--this process was iterated until $2000$ distinct parameter sets were generated. The $10$ Wigner $3j$-symbols with largest error, according to the metric $\epsilon_A:=|w_\mathrm{sym}-w_\mathrm{rec}|$ are shown, where $w_\mathrm{sym}$ and $w_\mathrm{rec}$ denote the value of a particular symbol as computed by the symbolic and recursive methods respectively; the average of these symbols is shown in red. The $1990$ symbols not shown, have a numerical error lower than that of the values displayed. In particular, we see that the numerical agreement between schemes is excellent (with machine epsilon $\varepsilon_m=2.2\times10^{-16}$) which implies an accurate implementation of the algorithm and stability for `large' parameter values. Note that the hybrid-recursive approach can be adapted (as demonstrated) to compute symbols with half-integral angular and projective quantum numbers.
	}
\end{figure}

In order to check the outlined algorithms we perform comparisons with a
symbolic calculation \cite{Olver:2010nistHandbook, Rasch:2003efficientStorage}.
Symbolic calculations are slow. Being of arbitrary accuracy they serve,
however,
as an excellent check. To verify our scheme all 
valid $\mathrm{SU}(2)$ $3j$-symbols were symbolically
calculated
up to a maximum entry of $j_1=10$. 
Figure~\ref{fig:wig3jerrorFig} shows the results of this procedure
-- excellent accuracy (machine precision) is achieved.

\section{(pseudo)-Spectral transformations: Tests and comparison}

In this section we describe a simple method for checking the associated
numerical error of our implementation of the spectral transformation outlined
in \S\ref{sec:specXform}. Further, we check that well-known 
exponential decay properties of
the magnitude of the representation of a function in spectral space are
satisfied. Although for the solution of a PDE, band-limits $L$ that probe the
border of stability of the TN recursive scheme (see \S\ref{sec:algodelCalc})
may be difficult to achieve, we future-proof our work by proposing a simple
linear/nonlinear hybrid scheme, in analogy to \S\ref{sec:CGw3jNumProc} that
restores stability.
We then briefly contrast a pseudo-spectral approach with
the full spectral scheme.

\subsection{Spectral transformation - Coefficient decay and \texorpdfstring{$\Delta$}{Delta}-stability}
\label{sec:errorStabilitySpeed}
As a preliminary check that the implementation of the spectral transformation
is sound we proceed as follows. First the generation of a complex-valued
function from a pseudo-random linear combination of the basis elements
${}_sY_{lm}(\vartheta,\varphi)$ is made. The random weights of the
aforementioned linear combination are denoted by $_s\tilde{a}_{lm}$. Taking the
inverse transformation of $_s\tilde{a}_{lm}$ allows for the construction of the
spatial representation of the function. Performing the forward transformation
in order to reconstruct the original random weights (reconstructed weights
denoted $_s\tilde{\alpha}_{lm}$) then allows for a comparison between
$_s\tilde{a}_{lm}$ and $_s\tilde{\alpha}_{lm}$. As we work with floating-point
arithmetic, we expect reconstruction to be within numerical tolerance. We
summarise this procedure as
\begin{equation}
  \nonumber
  {}_s\tilde{a}_{lm}
  \overset{\mathcal{F}{}^{-1}}{\longmapsto}
  \;{}_{s}f(\vartheta,\varphi)
  \overset{\mathcal{F}}{\longmapsto}\;{}_{s}\tilde{\alpha}_{lm}.
\end{equation}
By first separating the real and imaginary parts of each
${}_s\tilde{a}_{lm}$ as
\begin{equation}\nonumber
  {}_s\tilde{a}_{lm}=\Re{\left[{}_s\tilde{a}_{lm}\right]} + \i\Im{\left[{}_s\tilde{a}_{lm}\right]}
\end{equation}
we generate both $\Re{\left[_s\tilde{a}_{lm}\right]}$ and $\Im{\left[_s\tilde{a}_{lm}\right]}$ by sampling from the continuous uniform random distribution on the interval $[-1,1]$.

Although sampling random data as above provides us with a simple and robust diagnostic on our implementation, it is also informative to examine coefficient decay for specific test functions that are known to possess specific properties with respect to their spectral representation. 

\begin{figure}[ht]
\centering
\begin{subfigure}[Coefficient decay]{%
    \includegraphics[width=.48\linewidth] {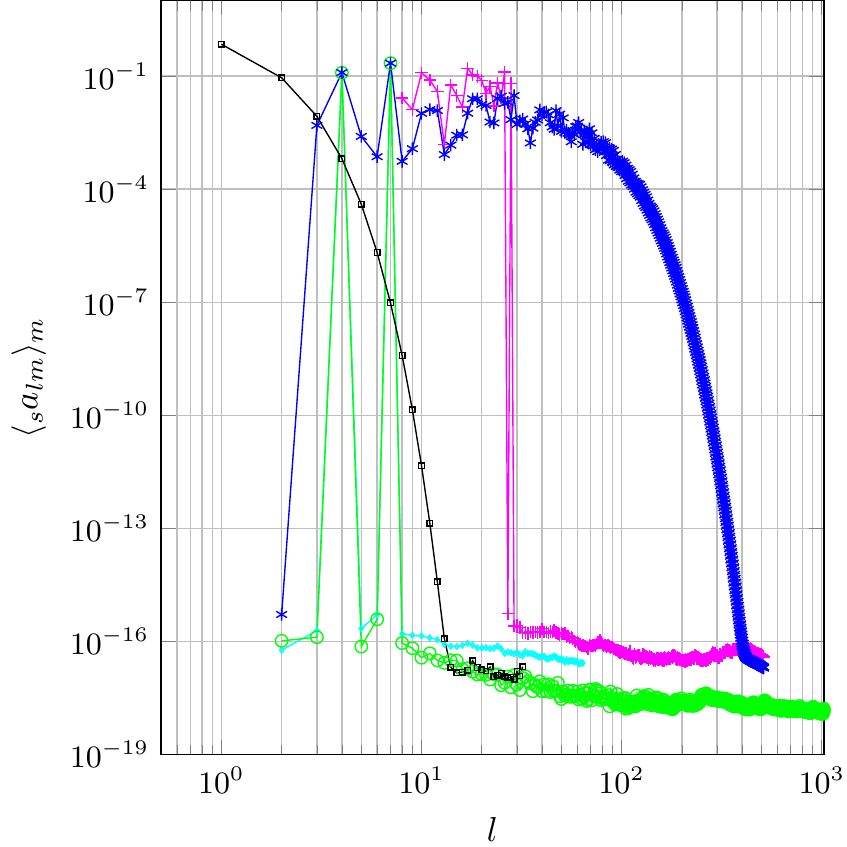}
   \label{fig:coeffDecay01}}%
\end{subfigure}
\hfill
 \begin{subfigure}[Fitted fall-off]{%
     \includegraphics[width=.48\linewidth]{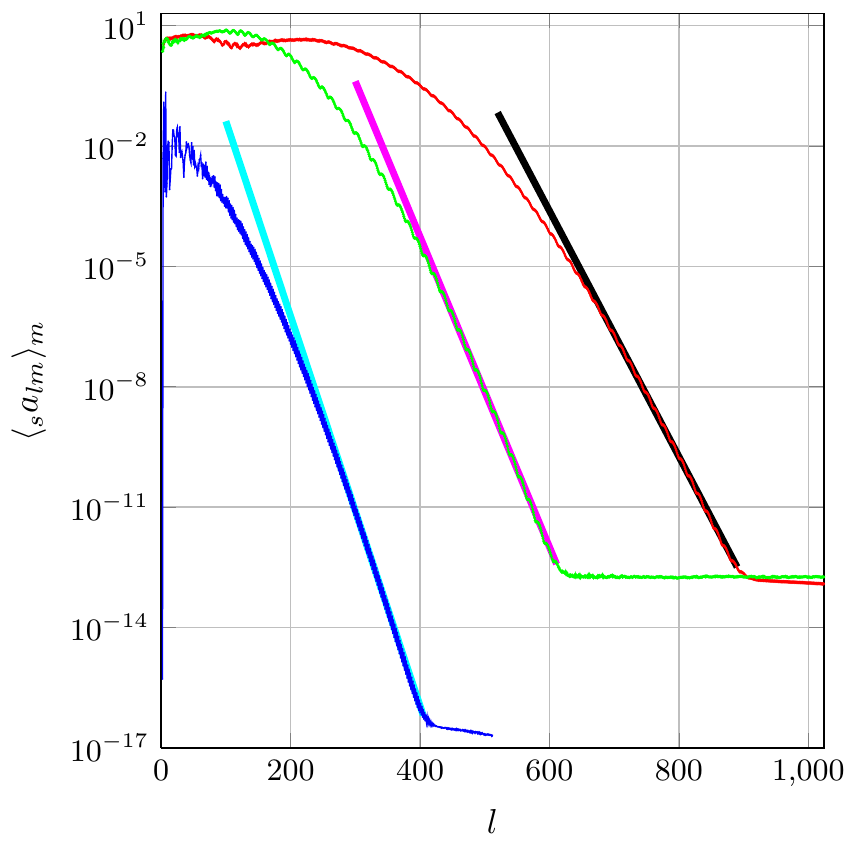}
     \label{fig:coeffDecay02}}%
\end{subfigure}%
\label{fig:coeffDecay}
\caption{ (Colour online)
a) Average magnitude of coefficients for fixed $l$ in the spectral representation of the functions in Eqs. (\ref{eq:testCoeffDec1}-\ref{eq:testCoeffDec4}).
	Cyan `$\diamond$': $s=2$ transformation of Eq.~\eqref{eq:testCoeffDec1} at a band-limit of $L=64$.
	Green `$\circ$': $s=2$ transformation of Eq.~\eqref{eq:testCoeffDec1} at a band-limit of $L=1024$.
	Magenta `$+$': $s=8$ transformation of Eq.~\eqref{eq:testCoeffDec2} at a band-limit of $L=512$.
	Blue `$*$': $s=2$ transformation of Eq.~\eqref{eq:testCoeffDec3} at a band-limit of $L=512$.
	Black `$\square$': $s=0$ transformation of Eq.~\eqref{eq:testCoeffDec4} at a band-limit of $L=32$.
	Note: The difference between coefficient magnitudes for transformations of Eq.~\eqref{eq:testCoeffDec1} at $L=64$ and $L=1024$ is solely due to the additional number of arithmetic operations required when performing a transform at a higher bandlimit $L$.
b) (Left to right) Blue `$*$': $s=2$ transformation of Eq.~\eqref{eq:testCoeffDec3} at a band-limit of $L=512$. Cyan line: linear fit of $l=(259,\,260,\,\dots,405)$, $\log(A_l)$.
Green `$\circ$': $s=-2$ transformation of Eq.~\eqref{eq:testCoeffDec5} at a band-limit of $L=1024$.
Magenta line: linear fit of $l=(408,\,409,\,\dots,611)$, $\log(A_l)$.
Red `$\diamond$': $s=-4$ transformation of Eq.~\eqref{eq:testCoeffDec6} at a band-limit of $L=1024$.
Black line: linear fit of $l=(660,\,661,\,\dots,890)$, $\log(A_l)$. See text for discussion.
	}
\end{figure}

We now inspect how the magnitude of coefficients of a function decay with increasing band-limit. In order to represent this decay in a convenient manner define the averaged coefficient:
\begin{equation}\label{eq:averagedCoeff}
A_l:=\langle{}_sa_{lm}\rangle_m := \sum_{m}\left|{}_s a_{lm}\right|/(2l+1).
\end{equation}
We now introduce the following smooth test functions:
\begin{align}\label{eq:testCoeffDec1}
{}_2 f(\vartheta,\varphi) &:= 1.1\,{}_2 Y_{4,1}(\vartheta,\varphi)-3.3\,{}_2 Y_{7,-6}(\vartheta,\varphi),\\\label{eq:testCoeffDec2}
{}_8 h(\vartheta,\varphi) &:= {}_2 f^4(\vartheta,\varphi),\\\label{eq:testCoeffDec3}
{}_2 g(\vartheta,\varphi) &:= {}_2 f(\vartheta,\varphi)\exp\left(-\left[-0.7\,{}_0 Y_{3,-2}(\vartheta,\varphi)+2.6\,{}_0 Y_{11,-9}(\vartheta,\varphi) \right]^2\right),\\\label{eq:testCoeffDec4}
{}_0 k(\vartheta,\varphi)&:= \exp\left(-{}_0Y_{1,-1}(\vartheta,\varphi)+{}_0Y_{1,1}(\vartheta,\varphi) \right),\\\label{eq:testCoeffDec5}
{}_{-2} q(\vartheta,\varphi) &:= \frac{1}{256}{}_{-2}Y_{25,-9}(\vartheta,\varphi)\exp\left(9.3\,{}_0Y_{20,-2}(\vartheta,\varphi)-22.5\,{}_0Y_{15,3}(\vartheta,\varphi) \right),\\\label{eq:testCoeffDec6}
{}_{-4} r(\vartheta,\varphi) &:= \frac{1}{4096} {}_{-2} q^2(\vartheta,\varphi)\exp\left(-5\,{}_0Y_{1,0}(\vartheta,\varphi) \right),
\end{align}
the spatial representation of which is initially constructed numerically.
Performing the forward transformation yields the spectral representation, which
together with Eq.~\eqref{eq:averagedCoeff} yields the averaged coefficient
decay which is shown in Fig.~\ref{fig:coeffDecay01}.
Equation~\eqref{eq:testCoeffDec1} and Eq.~\eqref{eq:testCoeffDec2} are clearly
comprised of a finite linear combination of SWSH and indeed this fact is
reflected by their complete capture 
at the band-limits tested. As
Eq.~\eqref{eq:testCoeffDec1} is completely captured at bandlimits $L=64$ and
$L=1024$, the only difference in the average magnitude of coefficients (Fig.~\ref{fig:coeffDecay01})
is solely due to the additional number of arithmetic operations that must be performed during the transformation.
Equation~\eqref{eq:testCoeffDec3} is a spin-$2$ function modulated in magnitude and phase by a smooth spin-$0$ function. In a similar manner Eq.~\eqref{eq:testCoeffDec5} and Eq.~\eqref{eq:testCoeffDec6} are smooth functions of spin weight $-2$ and $-4$ respectively. We expect that given smooth test functions the decay of the magnitude of the averaged coefficient $A_l$ should behave as $A_l\sim \alpha \exp(-\kappa l)$ $\alpha,\,\kappa\in\mathbb{R}$ for large $l$ \cite{katznelson2004harmonic ,boyd2001chebyshev}. This behaviour demonstrated in Fig.\ref{fig:coeffDecay02} where we construct a linear fit on a semilog plot, is characterstic of the expected exponential convergence that smooth functions should display \cite{boyd2001chebyshev}.

\subsection{Wigner-\texorpdfstring{$\Delta$}{Delta}-stability}\label{sec:wdelstab}
\begin{figure}[ht]
\centering
\begin{subfigure}[Trapani-Navaza]{\includegraphics[width=.49\linewidth] {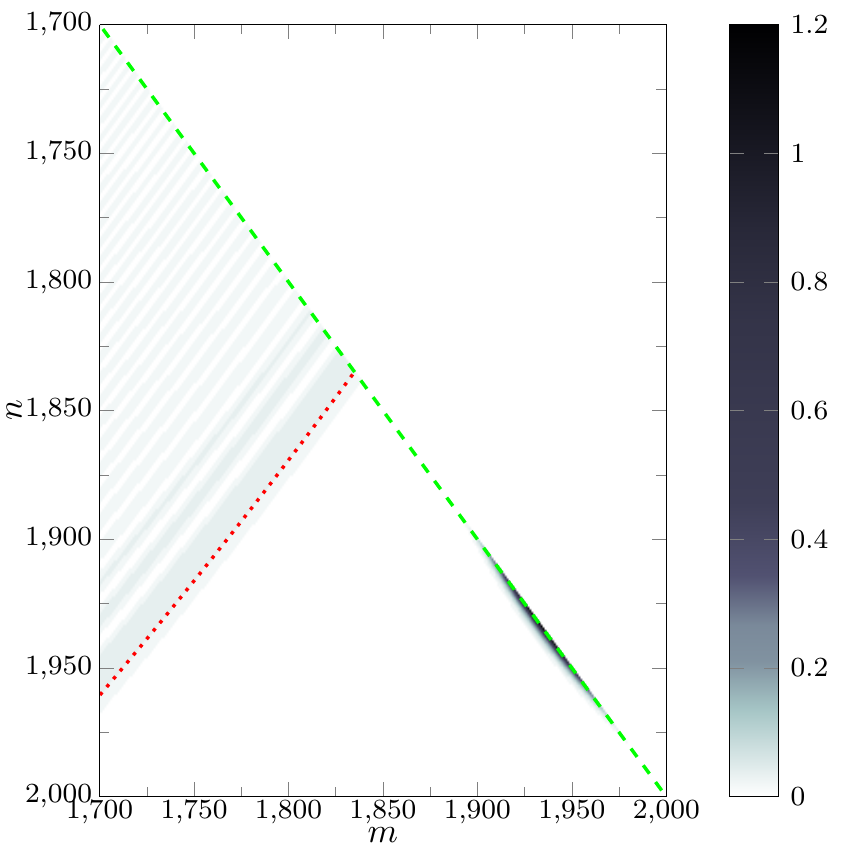}
   \label{fig:wignerDel_sub1}
 }%
\end{subfigure}\hfill
 \begin{subfigure}[Trapani-Navaza hybrid] {\includegraphics[width=.49\linewidth]{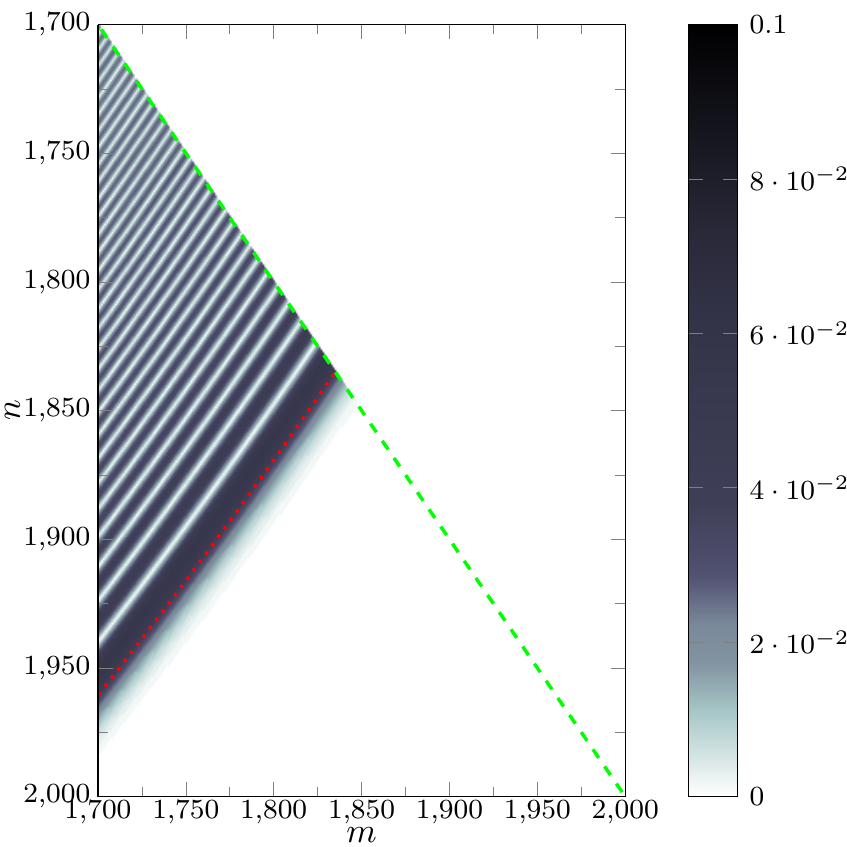}
   \label{fig:wignerDel_sub2}
 }%
\end{subfigure}%
\label{fig:wigDelHyb}
\caption{
(Colour online) Numerical instability of the Trapani-Navaza algorithm described in \S\ref{sec:algodelCalc} begins to occur for $L\simeq 2595$. a) $|\Delta^l_{mn}|$ under three-term linear recursion, with spurious values occurring in a region about $m\simeq1930,\,n\simeq1930$. It is clear that these values are incorrect as $|\Delta^l_{mn}|\leq 1$.
b) Hybrid scheme as discussed in text --- stability is maintained.
Dashed red arc indicates $\sqrt{|m|^2+|n|^2}$, dashed green line indicates edge of computational domain (prior to application of symmetries) for both recursion schemes.
Note: Color scales are distinct in both figures.
	}
\end{figure}

One aspect of the TN algorithm (briefly mentioned in \S\ref{sec:algodelCalc})
is that when transformations with a high band-limit are required instability
may arise. Figure ~\ref{fig:wignerDel_sub1} shows that instability occurs at
$L\simeq2595$; the situation is analogous to that when encountered whilst
performing calculations of the $3j$-symbols --- hence we proceed analogously.
In order to ameliorate the issue, we construct a new nonlinear recursion
relation.

Recall TN I-IV of \S\ref{sec:algodelCalc}. Fix $l$ and $m$. Observe that in the
region $m^2+n^2\geq l^2$ we are performing recursion with a three term linear
relation, furthermore, in this region we have values of $\Delta^l_{mn}$ that
monotonically increase in magnitude when $n$ is decreased from $l$. This leads
to instability of the TN algorithm. In order to avoid this, we 
apply Luscombe and Luban's general method
\cite{Luscombe:1998:SimplifiedW3jNL}, for the construction of a hybrid
linear/non-linear recursion to Trapani and Navaza's linear recursive scheme.  

To rewrite the three-term linear recursion relation stated in
Eq.~\eqref{eq:DelTN3tunst} as a two-term nonlinear recursion relation define
the ratio ${}_r\Delta^l_{mn}:=\Delta^l_{mn}/\Delta^l_{m(n-1)}$. This results in
\[
  {}_r\Delta^l_{m(n+1)}=\left(\frac{2m}{\sqrt{(l-n)(l+n+1)}}-{}_r\Delta^l_{m(n+2)}\sqrt{\frac{(l-n-1)(l+n+2)}{(l-n)(l+n+1)}}
  \right)^{-1}. \] To make use of this equation, we require a term
  $\Delta^l_{mn}$ within the range where nonlinear recursion is
  performed --- such that ratios may be converted to absolute values.
  While edge values with $n=l$ (for fixed $m$) may be used to
  calculate explicit terms from ratios, for reasons of numerical
  accuracy, it is advantageous to instead perform three-term iteration
  from $n=m$ to the edge of the region for which values are computed
  by the nonlinear scheme (see the dashed red arc in
  Fig.~\ref{fig:wignerDel_sub2}. Hence we now write a new three term
  relation for a fixed $n$ connecting three different values of $m$
  --- this is accomplished using the indicial symmetry given by the
  third of Eq.~\eqref{eq:DelIndSym} \[
  \Delta^l_{(m+2)n}=-\sqrt{\frac{(l-m)(l+m+1)}{(l-m-1)(l+m+2)}}\Delta^l_{mn}-\frac{2n}{\sqrt{(l-m-1)(l+m+2)}}\Delta^l_{(m+1)n}.
  \] 

\subsection{A comparison of pseudo-spectral and spectral schemes}\label{sec:PSandFScmp}

Consider two, smooth spin-weighted functions ${}_{s_1}f,\,{}_{s_2}g\in L^2(\mathbb{S}^2)$. Both ${}_{s_1}f$ and ${}_{s_2}g$ may be decomposed according to Eq.~\ref{eq:funcs2sphTrunc}, in terms of SWSHs of spin-weight $s_1$ and $s_2$ respectively, for some choice of band-limit $L$. As we will see in \S\ref{sec:advNum} product terms such as ${}_{s_1}f,\,{}_{s_2}g$ and hence products of the form ${}_{s_1}Y_{l_1m_1}\,{}_{s_2}Y_{l_2m_2}$ will need to be expanded in terms of a linear combination of ${}_{s_1+s_2}Y_{lm}$. We have already discussed how such expansions may be performed in a completely spectral scheme in \S\ref{sec:introCGw3j}, however when formulating the solution of non-linear PDEs in particular, multiple-sum terms can quickly become cumbersome without automatic code generation or some level of abstraction.

An alternative to the above is the pseudo-spectral method. Suppose we initially
have the coefficients ${}_{s_1}\tilde{f}_{l_1m_1}$ and
${}_{s_2}\tilde{g}_{l_2m_2}$ representing the functions ${}_{s_1}f$ and
${}_{s_2}g$ sampled (or intially seeded) at some band-limit $L$. The
coefficients ${}_{s_1+s_2}\tilde{a}_{lm}$ of the associated product
${}_{s_1}\tilde{f}\,{}_{s_2}\tilde{g}$ can be calculated by performing the
transformations:
\begin{align*}
\mathcal{F}^{-1}:&({}_{s_1}\tilde{f}_{l_1m_1})\mapsto {}_{s_1}\tilde{f}, 
& \mathcal{F}^{-1}:({}_{s_2}\tilde{g}_{l_2m_2})\mapsto {}_{s_2}\tilde{g};
\end{align*}
subsequently taking the pointwise product and transforming:
\begin{equation*}
\mathcal{F}^{-1}:{}_{s_1}\tilde{f}\,{}_{s_2}\tilde{g}
\mapsto ({}_{s_1+s_2}\tilde{a}_{lm}),
\end{equation*}
we find an approximation to an expansion utilizing Eq.\eqref{eq:prodSoln} directly. In order for the pseudo-spectral and spectral to coincide (to numerical tolerance) at some band-limit $L$ we have found that it is best to choose a band-limit for the pseudo-spectral scheme $\frac{2}{3}L_\mathrm{PS}\simeq L$, then truncating the constructed coefficients at a band-limit of $L$ -- this is the so-called `Orszag 2/3 rule' \cite{boyd2001chebyshev} which aids in the suppression of spurious aliasing.
We emphasize that this method also easily allows one to take into account the action of the $\eth,\,\eth'$ operators on functions by embedding their action as multiplication (see Eq.\eqref{eq:ethraco} and Eq.\eqref{eq:ethloco}) in coefficient space, together with taking account of their spin raising and lowering properties when transformations are performed.

\section{Numerical investigations of geometric initial value problems on the \texorpdfstring{$2$}{2}-sphere}\label{sec:NumInve}

Let $M:=\R\times\St$, and $\tau: M\rightarrow\R$ be the smooth
{time function} given by $\tau( t,p )=t$ for all $t\in\R$ and $p\in\St$ with non-vanishing gradient. For each
$p\in\St$, we consider the curve $\gamma_p:\R\rightarrow M$, $t\mapsto
(t,p)$, and the corresponding smooth tangent vector field $T=\gamma'_p$; in
particular we have $T(\tau)=1$. Because of this we can introduce coordinates $(t,\vartheta,\varphi)$ on
$M$, where $(\vartheta,\varphi)$ are standard polar coordinates on
$\St$ and where $T=\partial_t$. Let the level sets of $\tau$ be denoted by
\[\Sigma_t:=\{t\}\times\St\simeq \St, \quad t\in\R.\]
Any such subset is a smooth embedded submanifold of $M$. 
We make the same assumptions as in the last paragraph of
\Sectionref{sec:weightedquantities} -- in particular $U$ is a dense open subset
of $\St$ -- and we choose a smooth complex vector field\footnote{At this stage,
we make no further assumptions about $m$; in particular, $m$ should not be
confused with the field in \Eqref{eq:mmStar}. We shall choose $m$ specifically
in the two applications discussed below. Notice also that since there is no
metric defined on $M$ so far, we do also not assume a normalization for $m$
yet.}
$m$ on $\R\times U$ which is tangent to $\Sigma_t$ at each $t$ and which
satisfies $m(\tau)=0$. Let us suppose that $(b_0,b_1,b_2):=(T,m,\mbar)$ is a
smooth frame almost everywhere on $M$. We let $(\alpha^0,\alpha^1,\alpha^2)$ be
the corresponding dual frame and hence deduce that $\alpha^0=d\tau$. The
following notion of {spin-weight} based on frame transformations of the form
\begin{equation}\nonumber
T\mapsto T, \quad m\mapsto e^{\i\rho} m,
\end{equation}
where $\rho$ can be an arbitrary smooth function on $M$, is useful for the
following discussion. Any quantity $h$ on $M$, which behaves likes $h\mapsto
e^{\i s\rho} h$ under this transformation, is said to have spin-weight $s$. For
instance, the frame vector $T$ has spin-weight $0$, $m$ has spin-weight $1$ and
$\mbar$ has spin-weight $-1$.

Of particular importance for the following discussions are commutators of the frame fields
\[
\commutator ijk:=\pairing{\alpha^i}{[b_j,b_k]}.
\]
The assumptions above yield
\[
\commutator 012=0,
\]
and all commutators can be computed from the following ones
\begin{equation}\nonumber
  \commutator 001=:\kappa_1,\quad
  \commutator 201=:\kappa_2,\quad
  \commutator 101=:\mu_0+L_0,\quad
  \commutator 212=:\mu_1+L_1.
\end{equation}
The functions $\kappa_1$, $\kappa_2$, $\mu_0$, $\mu_1$, $L_0$ and $L_1$ have the following transformation behavior under the frame transformations above:
\[
\kappa_s\mapsto e^{\i s \rho}\kappa_s,\quad \mu_s\mapsto e^{\i s \rho}\mu_s,\quad\text{for $s=0,1,2$},
\]
i.e.,\ these are functions have a well-defined spin-weight $s$,
while
\begin{equation}\nonumber
  L_0 = \i T(\rho),\quad L_1 = -\i m(\rho) e^{\i\rho},
\end{equation}
and hence do not have a spin-weight. Notice that in particular, $L_0=L_1=0$ for $\rho\equiv 0$, i.e., for the reference frame.

\subsection{Tensor advection equation}\label{sec:advProb}

We want to start with the following \highlight{advection problem}. Let $V$ be a
given smooth vector field on $M$ with $V(\tau)=1$. Pick a smooth
$(r,0)$-tensor field $N_*$ 
in a neighborhood
of $\Sigma_0$. We consider the
initial value problem
\begin{equation}
  \label{eq:advectionproblem}
  \lie{V}{N}=0,\quad \left.N\right|_{t=0}=\left.N_*\right|_{t=0},
\end{equation}
for an unknown $(r,0)$-tensor field $N$ on $M$.

Since $V$ is smooth it generates a flow on $M$ which maps each point $p$ of $M$
to a point on the integral curve of $V$ through $p$. Due to the condition
$V(\tau)=1$, it follows that $V$ has a non-vanishing `time component' in the
direction of $T$ and, in general, a non-vanishing `spatial component'
tangential to $\Sigma_t$. Solutions $N$ of \eqref{eq:advectionproblem} are
invariant under this flow generated by $V$ and are therefore advected in time
along the `spatial component' of $V$. We notice that in the case $r=0$,
\eqref{eq:advectionproblem} reduces to the standard scalar advection equation.
Moreover, we observe that this initial value problem does not require the
specification of a metric on $M$.

In order to bring \Eqref{eq:advectionproblem} into a form for which our
formalism applies, we choose the frame $(b_0,b_1,b_2)= (T,m,\mbar)$ above with
$T=\partial_t$ and $m=m^*$ given by \Eqref{eq:standardreferenceframe}. Hence
the dual frame is $(\alpha^0,\alpha^1,\alpha^2)=(n,\sigma,\bar\sigma)$ with
\[n=d\tau,\quad \sigma=\frac 1{\sqrt 2}(d\vartheta + \i \sin\vartheta d\varphi).\]
With this it follows that
\begin{equation}
  \label{eq:commadvect}
  \kappa_1=\kappa_2=\mu_0=0,\quad \mu_1=-\frac 1{2\sqrt 2}\cot\vartheta.
\end{equation}
Since the vector field $V$ has the property $V(\tau)=1$, there exists a smooth complex function $\xi$
so that
\begin{equation}
  \label{eq:decompV}
  V=T+\sqrt 2\, \xi m+\sqrt 2\,\bar\xi\,\mbar.
\end{equation}
Notice that $\xi$ has spin-weight $-1$ (and hence $\bar\xi$
spin-weight $1$). For simplicity, we assume that $\xi$ is independent of $t$.

We restrict to the scalar and vectorial advection problems now, i.e.\ to $r=0$ and $r=1$. In the scalar case, $N$ is a
function with spin-weight zero hence Eq.~\eqref{eq:advectionproblem}
translates to
\begin{equation}
  \label{eq:advectionproblemscalar}
  \partial_t N=-\xi\, \eth N-\bar\xi\,\etp N,\quad \left.N\right|_{t=0}=\left.N_*\right|_{t=0},
\end{equation}
using \eqref{eq:ethm} and \eqref{eq:ethm2} for $s=0$.
We check easily that each term in this equation is of spin-weight
$0$. In general, consistency of the spin-weights of the terms in
an equation
is a good indication that it has been derived correctly.

In the vectorial case, we decompose the vector $N$ as
\begin{equation}
   \label{eq:decompN}
  N={}_{0} N\, T+\sqrt 2\, {}_{-1} N\, m+\sqrt 2\, {}_{1} N\, \mbar,
\end{equation}
where ${}_{s} N$ is of spin-weight $s$ and where ${}_{1}
N=\overline{{}_{-1} N}$. Projecting \eqref{eq:advectionproblem} onto $n$, $\sigma$ and $\bar\sigma$, and using \eqref{eq:commadvect}, \eqref{eq:decompV} and \eqref{eq:decompN} and the fact that $L_0$ and $L_1$ vanish for the reference frame, we find
\[
\begin{aligned}
  \partial_t ({}_{0} N)  & +\sqrt 2\,\xi m({}_{0} N)+\sqrt 2\,\bar\xi\, \mbar({}_{0} N)=0,\\
  \partial_t ({}_{-1} N) & +\sqrt 2\,\xi m({}_{-1} N)+\sqrt 2\,\bar\xi \mbar({}_{-1} N)
- \sqrt 2\,{}_{-1} N m(\xi) \\
& -\sqrt 2\,{}_{1} N\mbar(\xi) +{}_{1} N\xi \cot\vartheta
-{}_{-1} N\bar\xi\cot\vartheta=0,\\
\partial_t ({}_{1} N) & +\sqrt 2\,\bar\xi \mbar({}_{1} N)+\sqrt 2\,\xi m({}_{1} N)
-\sqrt 2\,{}_{1} N\mbar(\bar\xi)\\ & -\sqrt 2\,{}_{-1} N m(\bar\xi) +{}_{-1} N\bar\xi \cot\vartheta
-{}_{1} N\xi\cot\vartheta=0.
\end{aligned}
\]
Using the relations \eqref{eq:ethm} and \eqref{eq:ethm2} to write
\[m({}_s f)=\frac{1}{\sqrt 2}\left(\eth({}_s f)+ {}_s f s\cot\vartheta\right),\quad
\mbar({}_s f)=\frac{1}{\sqrt 2}\left(\etp({}_s f)-{}_s f s\cot\vartheta\right),\]
for any quantity ${}_s f$ of spin-weight $s$, we find
\begin{equation}
\label{eq:advectionproblemvector}
	\begin{aligned}
		\partial_t ({}_{0} N)&=-\xi\, \eth({}_{0} N)-\bar\xi\, \etp({}_{0} N),\\
		\partial_t ({}_{-1} N)&=-\xi\, \eth({}_{-1} N)-\bar\xi\, \etp({}_{-1} N)
		+\eth(\xi)\, {}_{-1} N +\etp(\xi)\, {}_{1} N,\\
		\partial_t ({}_{1} N)&=-\bar\xi\, \etp({}_{1} N)-\xi\, \eth({}_{1} N)
		+\etp(\bar\xi)\, {}_{1} N +\eth(\bar\xi)\, {}_{-1} N.
	\end{aligned}
\end{equation}
We realize that all terms are of a well-defined and consistent spin-weight and that all formally singular terms (i.e., those proportional to $\cot\vartheta$) disappear (as one expects). The first of Eqs.~\eqref{eq:advectionproblemvector} is of the same form as \eqref{eq:advectionproblemscalar} and is decoupled from the other two. The third equation is the complex conjugate of the second one.

\subsubsection{Spectral decomposition and numerical results}\label{sec:advNum}

In this section we solve~\eqref{eq:advectionproblemscalar} and~\eqref{eq:advectionproblemvector} together with appropriate initial data by application of the Fourier-Galerkin method. We will choose advecting fields $\{{}_{-1}\xi:=\xi,\,{}_{1}\xi:=\bar\xi\}$ comprised of linear combinations of axial rotations. This choice is particularly amenable to analysis of numerical error and stability since initial data that is advected by such fields will undergo a time evolution that must periodically return to its initial state --- it is this periodic behaviour that we exploit for our numerical tests.

Recall that the generators of rotations in $\R^3$
are given by:
\[
X = -\sin\varphi\,\partial_\vartheta-\cot\vartheta\cos\varphi\,\partial_\varphi,\quad
Y = \cos\varphi\,\partial_\vartheta-\cot\vartheta\sin\varphi\,\partial_\varphi,\quad
Z = \partial_\varphi,
\]
respectively. If we choose $V$ to be one of these generators and then
decompose $V$ as in \eqref{eq:decompV} we find 
\begin{align*}
	{}_{-1}\xi_X&=-\i\sqrt{\frac{2\pi}{3}}\left({}_{-1}Y_{1,-1}-{}_{-1}Y_{1,1} \right),
	&{}_{1}\xi_X&=-\i\sqrt{\frac{2\pi}{3}}\left({}_{1}Y_{1,-1}-{}_{1}Y_{1,1} \right);\\
	{}_{-1}\xi_Y&=\sqrt{\frac{2\pi}{3}}\left({}_{-1}Y_{1,-1}+{}_{-1}Y_{1,1} \right),
	&{}_{1}\xi_Y&=\sqrt{\frac{2\pi}{3}}\left({}_{-1}Y_{1,-1}+{}_{-1}Y_{1,1} \right);\\
	{}_{-1}\xi_Z&=-2\i\sqrt{\frac{\pi}{3}}\,{}_{-1}Y_{1,0},
	&{}_{1}\xi_Z&=-2\i\sqrt{\frac{\pi}{3}}\,{}_{1}Y_{1,0}.
\end{align*}
It is convenient to further normalize the advecting fields ${}_{\pm1}\hat{\xi}_{(.)}:=\sqrt{2}\pi\,{}_{\pm1}\xi_{(.)}$
such that the final state of the fields being advected will again coincide with the choice of the initial field configuration after a temporal period $\mathcal{T}=1$.

For general data Eq.~\eqref{eq:funcs2sphTrunc} gives:
\begin{equation}\label{eq:advExpa}
  \begin{aligned}
{}_{s}\xi(\vartheta,\varphi)
&=
\sum_{l=|s|}^L\sum_{m=-l}^{l}\,{}_{s}\xi_{lm}\,{}_{s} Y_{{l}{m}}(\vartheta,\varphi),\\
{}_{s} N(\vartheta,\varphi;t)
&=
\sum_{l=|s|}^L\sum_{m=-l}^{l}\,{}_{s}N_{lm}(t)\,{}_{s} Y_{{l}{m}}(\vartheta,\varphi),
\end{aligned}
\end{equation}
for advecting fields and the fields being advected respectively. The
time-dependence of the solution is carried by the expansion
coefficients ${}_{s}N_{lm}(t)$.
Equations (\ref{eq:ethraco}, \ref{eq:ethloco}, \ref{eq:sYlmOrth}, \ref{eq:SpinnySpectroscope}, \ref{eq:advectionproblemscalar}, \ref{eq:advExpa}) lead to the spectral representation of the scalar advection problem:
\begin{equation}
  \label{eq:advScSpec}
  \begin{aligned}
    {}_0\dot{N}_{lm}(t)
    &=
    \sum^L_{l_a,l_b=1}\sum^{l_a}_{m_a=-l_a}
    \left[\delta_{l_b\geq|m-m_a|}\right]
    \left\{\vphantom{\sum^L_{l_a=1}}
      \sqrt{l_b(l_b+1)}
      \mathcal{A}_l(-1,l_a,m_a;\,1,l_b,m-m_a)
    \right.\\
 &
\left.\times{}_{0}N_{l_b,m-ma}(t)
	\left({}_{-1}\xi_{l_a,m_a}-(-1)^{l_a+l_b+l}{}_{1}\xi_{l_a,m_a)} 
	\right)
\vphantom{\sum^L_{l_a=0}}\right\},
\end{aligned}
\end{equation}
where $\left[\delta_{l_b\geq|m-m_a|}\right]$ is a Boolean function -- equal to $1$ when the inequality is satisfied and $0$ when it is not. We introduce such functions to make explicit the distinct limits of summation of the terms in Eq.~\eqref{eq:advScSpec}.

For the vector advection problem the spectral representation of the spin $0$ component of the system is again given by Eq.~\eqref{eq:advScSpec}. Due to the condition ${}_{1}N=\overline{{}_{-1}N}$ and Eq.~\eqref{eq:yslmconj} we have ${}_{1}N_{lm}=\overline{{}_{-1}N_{l,-m}}$, hence in order to completely specify the system, only the spin $-1$ equation is required:
\[
\begin{aligned}
	{}_{-1}\dot{N}_{lm}(t)
	=&
	\sum_{l_a=1}^L\sum_{m_a=-l_a}^{l_a}\sum_{l_b=1}^L
	\left[\delta_{l_b\geq|m-m_a|}\right]
	\left\{\vphantom{\sum_{l_a=1}^L}
		\sqrt{l_b(l_b+1)}
		\left(
			{}_{-1}\xi_{l_a,m_a}\,{}_{-1}N_{l_b,m-m_a}(t)
		\right.
	\right.\\
	 &
		\left.-{}_{-1}\xi_{l_b,m-m_a}\,{}_{-1}N_{l_a,m_a}(t)\right)
		\mathcal{A}_l(-1,l_a,m_a;\,0,l_b,m-m_a)\\
	 &
	+\sqrt{(l_b-1)(l_b+2)}
	\left(
		{}_{-1}\xi_{l_b,m-m_a}{}_{1}N_{l_a,m-m_a}(t)
	\right.\\
	&
	\left.
		-{}_{1}\xi_{l_a,m-m_a}{}_{-1}N_{l_b,m-m_a}(t)
	\right)\mathcal{A}_l(1,l_a,m_a;-2,l_b,m-m_a)\left[\delta_{l_b\geq 2}\right]
	\left.\vphantom{\sum_{l_a=1}^L}\right\}
\end{aligned}
\]

\begin{figure}[ht] 
	\centering
	\includegraphics{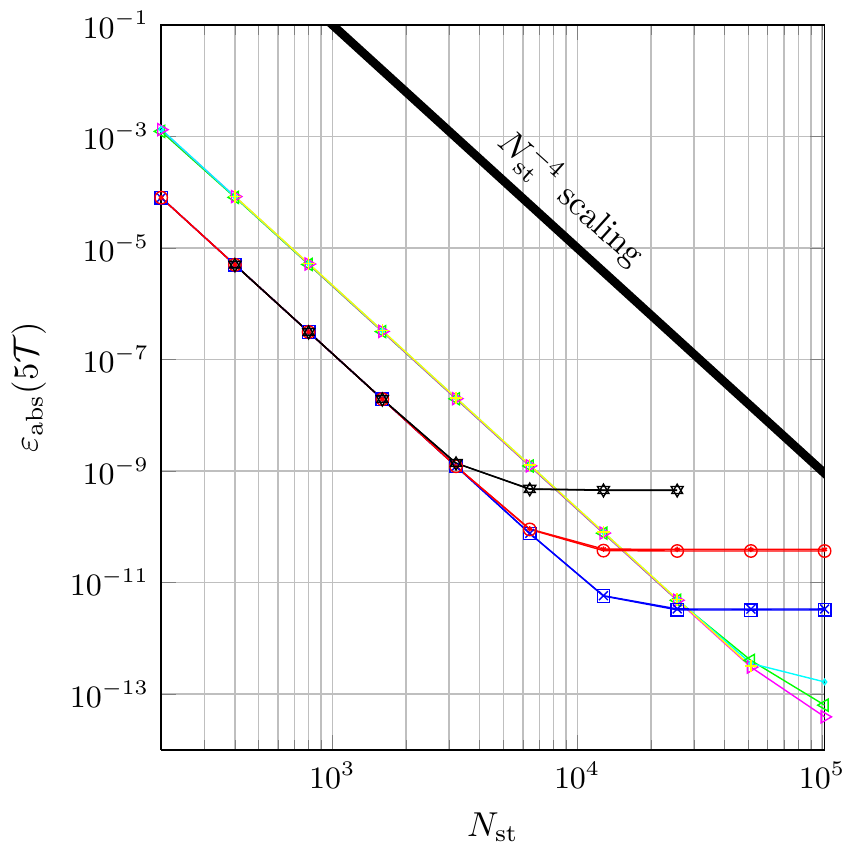}
	\caption{
	(Colour online) Absolute difference associated with initial condition and stroboscopically sampled evolution
	under specified advecting field after $5$ periods elapse.
	(\emph{Advection of scalar field}):
	 Green `$<$': $L=4$; Magenta `$>$': $L=8$; Cyan `$\diamond$': $L=16$; Yellow `$+$': $L=32$: advecting field corresponds to a $z$ axial rotation, initial data corresponds
	to Eq.~\eqref{eq:testCoeffDec4}.
	(\emph{Advection of vector field}):
	Blue: $L=8$, `$\times$': $s=-1$ component, `$\square$': $s=1$ component;
	Red: $L=16$, `$*$': $s=-1$ component, `$\circ$': $s=1$ component;
	Black: $L=32$, `$\triangledown$': $s=-1$ component, `$\triangle$': $s=1$ component:
	Advecting field corresponds to the normalized sum of $x$ and $y$ axial rotations, initial data corresponds to a $z$ axial rotation. We find excellent agreement with the expected $4$th order convergence, prior to a saturation in convergence due to numerical roundoff error.
	}
	\label{fig:advectionSweepNst}
\end{figure}

Figure~\ref{fig:advectionSweepNst} shows the results of numerical convergence
tests performed using an explicit $4$th order Runge-Kutta (RK) method in time for a variety
of time-steps $N_\mathrm{st}$ and band-limits $L$.
We show the absolute value of the maximum difference in coefficient
space $\varepsilon_\mathrm{abs}$ between initial configurations of test fields
and their final configurations upon advection for $5$ periods $\mathcal{T}$.
Excellent agreement with the expected $4$th order convergence is displayed. We
find that in both scalar and vector cases the error associated with the
temporal discretization dominates that of the spatial scheme -- this was verified
by constructing a semilog plot of data generated for various values of $L$ vs.
$\varepsilon_\mathrm{abs}(5\mathcal{T})$ at fixed $N_\mathrm{st}$
where we found a horizontal line.
This is not unexpected as we are advecting smooth fields, by
smooth fields and expect exponential convergence for the spatial sampling (see
\S\ref{sec:errorStabilitySpeed}) whereas only $4$th order convergence in time
is provided by the Runge-Kutta method. Note, that here and in what follows when performing numerical expansions as in Eqs.~(\ref{eq:ClebschGordanSeries},\ref{eq:prodSoln}) all terms with $l>L$ are discarded.

\subsection{The \texorpdfstring{$2+1$}{2+1}-vacuum Maxwell equations on the \texorpdfstring{$2$}{2}-sphere}\label{sec:maxCnstrct}
As a second application, we study the initial value problem of the 
$2+1$-vacuum Maxwell equations on $M$, i.e., the equations
\[dF=0,\quad \delta F=*d*F=0,\]
for the electromagnetic $2$-form $F$ on $M$, where $d$ is the exterior
derivative and $*$ is the Hodge dual associated with a Lorentzian metric $g$ on
$M$; we assume the signature $(+,-,-)$. In abstract index notation, these
equations can also be written as
\[\nabla_{[\mu} F_{\nu\rho]}=0,\quad \nabla^\mu F_{\mu\nu}=0.\]

Let $(e_0,e_1,e_2)$ be an orthonormal frame with respect to $g$ 
and choose the frame $(b_0,b_1,b_2)=(T,m,\bar m)$
by $e_0=b_0=T$ and $m=(e_1 - \i e_2)/\sqrt 2$. This implies for example
$g(T,T)=1$, $g(m,\bar m)=-1$ and $g(m,m)=0$.
Let $(\omega^0,\omega^1,\omega^2)$ be the coframe dual to $(e_0,e_1,e_2)$. 
If $(\alpha^0,\alpha^1,\alpha^2)=(n,\sigma,\bar\sigma)$ is the coframe dual to
$(T,m,\bar m)$ then
\[\omega^0=n=d\tau,\quad \sigma=\frac 1{\sqrt 2}(\omega^1 + \i\omega^2).\]
Since any smooth metric on $\St$ is conformal to the standard metric on the unit sphere,
a general smooth metric on $M$  is described by a smooth real\footnote{We could also assume that $f$ is complex. Its phase, however, would generate nothing but a rotation of the frame and would hence not contribute to the metric.} strictly positive function $f:M\rightarrow\R$ of spin-weight zero so that
\[m=f m^*,\]
where $m^*$ is given by \Eqref{eq:standardreferenceframe}.
In the special case $f\equiv 1$, one obtains the geometry of the standard round unit sphere.
Physically, our system can therefore be interpreted as vacuum electrodynamics in a  universe of $2$-dimensional spatial spherical topology whose geometry is described by the function $f$. Notice that we allow $f$ to depend on time.
As before, we assume $T=\partial_t$. Then, it follows
\[\kappa_1=\kappa_2=0,\quad \mu_0=\frac{\partial_t f}{f},\quad \mu_1=m^*(f)-\frac 1{\sqrt 2}f\cot\vartheta.\]

In $2+1$ dimensions, the $2$-form $F$ has three independent components which we write as
\[F=E_1\omega^1\wedge\omega^0+E_2\omega^2\wedge\omega^0+B \omega^1\wedge\omega^2.\]
Hence $E_1$ and $E_2$ can be interpreted as the two components of the electric
field and $B$ as the component of the magnetic field. In fact, we can
introduce a purely spatial ``electric'' one-form
\[E=-E_1\omega^2+E_2\omega^1,\]
so that the one-form $*F$ becomes
\[*F=E-B \omega^0.\]
For the following it is useful to define the three complex Maxwell scalars as
\begin{align*}
{}_{1}F&:=F(T,m)=-\frac 1{\sqrt 2}(E_1 - \i E_2),\quad {}_{0}F:=\i F(m,\mbar)=-B,\\ {}_{-1}F&:=F(T,\mbar)=-\frac 1{\sqrt 2}(E_1 + \i E_2),
\end{align*} 
where ${}_{s}F$ has spin-weight $s$. These fields satisfy the reality
conditions
\begin{equation}\label{eq:FreCond}
{}_{0}F=\overline{{}_{0}F},\quad {}_{-1}F=\overline{{}_{1}F}.
\end{equation}
We find that the electric one-form $E$ can be written as
\begin{equation}\label{eq:EfieldspComb}
E = -\i \,{}_{1}F\sigma + \i \,{}_{-1}F\bar\sigma.
\end{equation}

The first Maxwell equation $dF=0$ is equivalent to $\delta *F=0$, and hence to
\[\delta (B n)=\delta E.\]
This equation corresponds symbolically to the Maxwell equation ``$\dot
B=\text{div E}$''. Using the same arguments as for the advection problem,
we arrive at the following evolution equation for the magnetic field
\begin{equation}\label{eq:maxEv0}
\partial_t\,{}_{0}F
  =\frac{\i}{\sqrt 2} \left(\,{}_{1}F \eth' f-\,{}_{-1}F\eth f+f \eth {}_{-1}F-f \eth' {}_{1}F\right)
  +2 \,{}_{0}F \frac{\partial_t f}f.
\end{equation}
The second Maxwell equation $\delta F=0$ is equivalent to $d *F=0$, and hence to
\[*d (B n)=*d E.\]
This corresponds to three equations. The $n$-component is the constraint
\begin{equation}
\label{eq:constraint} 
0=(*dE)(T)=\frac{1}{\sqrt 2}\left(\eth f\,{}_{-1}F-f\eth\,{}_{-1}F 
+ \eth' f \,{}_{1}F-f\eth' {}_{1}F\right)=:\frac{C}{\sqrt 2}.
\end{equation}
The $\sigma$- and $\bar\sigma$-components yield evolution equations for the components of the electric field
\begin{align}\label{eq:maxEv1}
\partial_t\,{}_{-1}F
  &=
  -\frac 1{\sqrt 2} \i\, f\, \eth'\,{}_{0} F+\frac{\partial_t{ f}}{ f}\, {}_{-1}F,\\\label{eq:maxEv2}
\partial_t\,{}_{1}F
  &=
  \frac 1{\sqrt 2} \i\, f\, \eth\,{}_{0}F+\frac{\partial_t f}f\, {}_{1}F.
\end{align}

We see explicitly that all terms in the equations are non-singular and of consistent spin-weight. Moreover, when we write the $\eth$-operator in a coordinate basis, we see that the evolution system is symmetric hyperbolic and hence gives rise to a well-posed initial value problem. It remains to show that the constraint $C=0$ propagates under the evolution. For this we derive the evolution equation for the constraint violation quantity $C$ given in \Eqref{eq:constraint}. We take the time derivative of \Eqref{eq:constraint} and use the evolution equations of ${}_{-1}F$, ${}_{0}F$, ${}_{1}F$.
Then it is straightforward to find that
\begin{equation}\label{eq:constrEvF}
\partial_t C=2\frac{\partial_t f}f C.
\end{equation}
It follows that if the initial data satisfy the constraint, i.e., $C=0$ at the
initial time, then $C\equiv 0$ for all times and hence the constraints will be
satisfied, up to machine precision, during the whole evolution.
Furthermore, as can easily be seen from the evolution equations, the reality conditions of Eq.~\eqref{eq:FreCond} are preserved during the whole evolution provided they are fulfilled at the initial time.

Notice that $0=(*dE)(T)$ is equivalent to
$dE(e_1,e_2)=0$. 
Since $E$ is purely spatial
this is equivalent to $\tilde
dE=0$, where $\tilde d$ is the purely spatial exterior derivative on the
initial hyper-surface. Because
$\St$ is simply connected it follows that every solution $E$ of the
constraint is of the form
\begin{equation}\label{eq:EfieldPot}
E=\tilde d\Phi=\frac{f}{\sqrt{2}}\left\{(\eth \Phi)\sigma+(\eth'\Phi)\overline{\sigma} \right\},
\end{equation}
where $\Phi$ is an arbitrary smooth scalar function of spin-weight zero on the initial hyper-surface. Comparison of Eq.~\eqref{eq:EfieldPot} with Eq.~\eqref{eq:EfieldspComb} yields
\begin{equation}\label{eq:iniCondMaPot}
{}_{-1}F = -\i \frac{f}{\sqrt{2}}\eth' \Phi, \quad {}_{1}F = \i \frac{f}{\sqrt{2}}\eth \Phi.
\end{equation}
As the constraint Eq.~\eqref{eq:constraint} is independent of ${}_{0}F=-B$
the magnetic field may be prescribed freely, subject only to reality
conditions.

\subsubsection{Numerical results}\label{sec:numMax}
Our goal is now the construction of numerical solutions by means of spectral
decompositions of the dynamical equations governing the Maxwell system. In
order to check the numerical solutions thus constructed, we examine 
the preservation of
the constraints associated with the system. In order to further ensure that our
implementation is accurate and robust we compare our
calculation to a pseudo-spectral implementation based on discussion of
\S\ref{sec:PSandFScmp}.
We proceed in two stages: We consider the case where $\mathbb{S}^2$ is
deformed at the initial time and examine how a simple choice of initial
data evolves with this fixed geometry.
We then allow for a time-dependent change of the geometry, again examining
how our solution for the fields develops with time.

As the method is entirely analogous to that of \S\ref{sec:advNum} we do not
explicitly state our expansions of
Eqs.~(\ref{eq:maxEv0}, \ref{eq:maxEv1}, \ref{eq:maxEv2}, \ref{eq:iniCondMaPot},
\ref{eq:constraint})
here. However it is worth pointing out that due to~\eqref{eq:FreCond}
and~\eqref{eq:yslmconj} we again have
\[
{}_{0}F_{l,m}=\overline{{}_{0}F_{l,-m}}(-1)^m,\quad
{}_{1}F_{l,m}=\overline{{}_{-1}F_{l,-m}}(-1)^{1-m},
\]
which implies that only a subset of the full dynamical system need be evolved, the rest may be extracted by these symmetries. We have also found it convenient to re-expand terms such as $(\partial_tf)/f$ by defining an auxiliary function:
\[
g(\vartheta,\varphi;t):=\partial_t\left(\ln(f(\vartheta,\varphi;t))\right) =
\frac{\partial_t f}{f}
=
\sum_{l=0}^L\sum_{m=-l}^{l}\,{}_{0}\beta_{l,m}(t)\,{}_{s} Y_{l,m}(\vartheta,\varphi),
\]
rather than dealing with $f$ directly.

\begin{figure}[ht]
\centering
\begin{subfigure}[Spectral method]{
    \includegraphics[width=.48\linewidth] {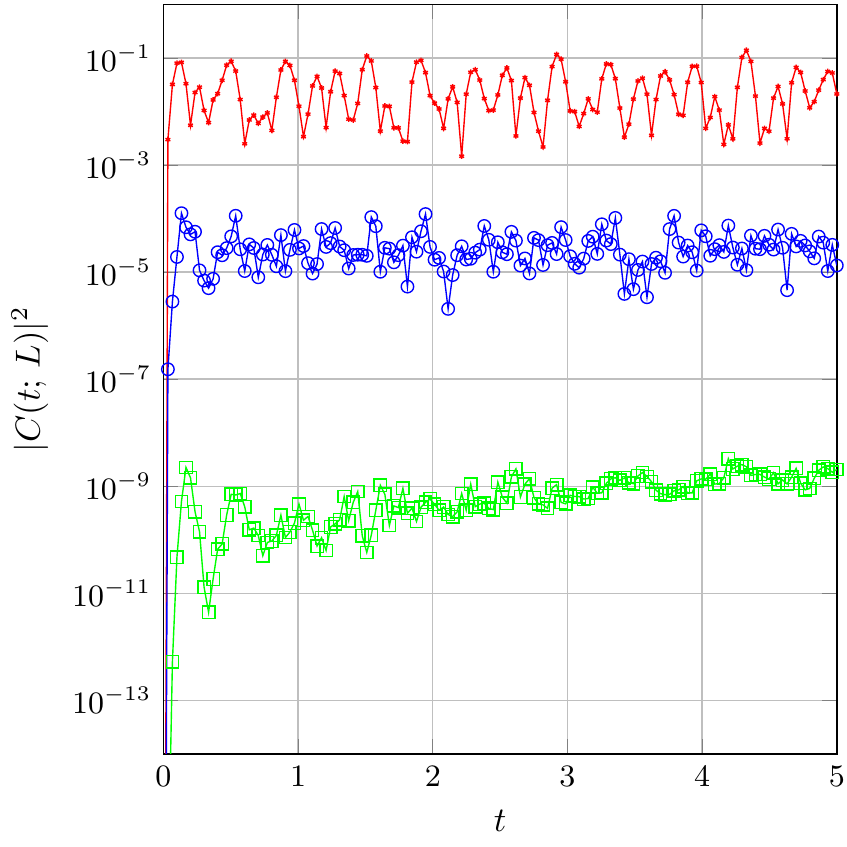}
    \label{fig:constr_sub1}}%
\end{subfigure}
\hfill 
\begin{subfigure}[Pseudo-spectral method]
{\includegraphics[width=.48\linewidth]{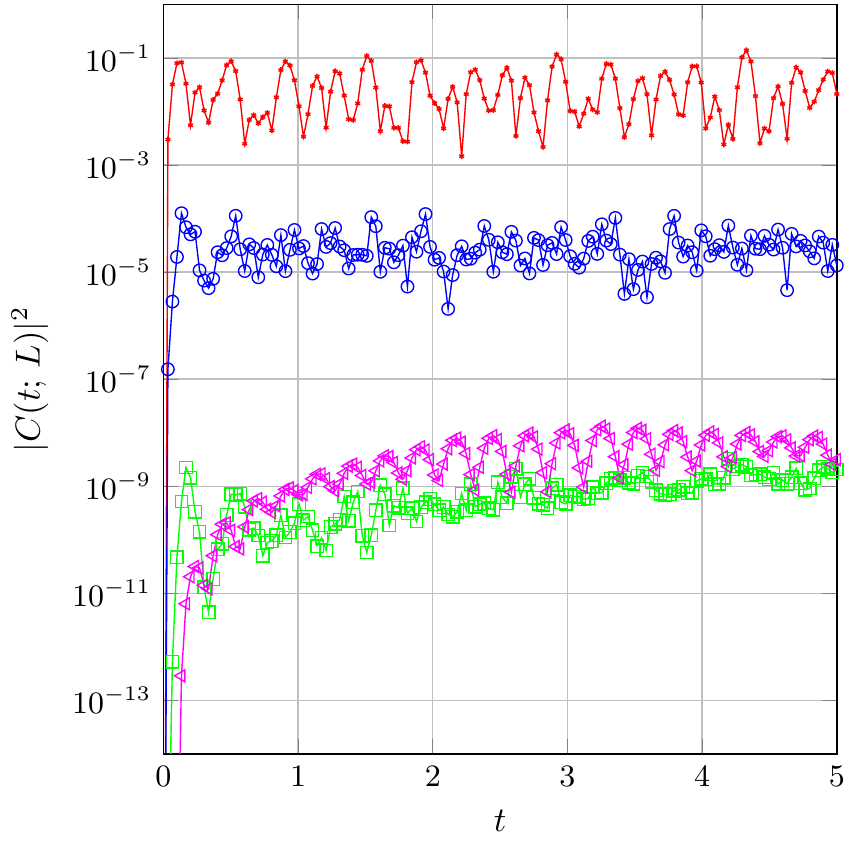}
   \label{fig:constr_sub2}}%
\end{subfigure}%
\caption{\label{fig:constr}
(Colour online)
Value of the constraints (Eq.~\eqref{eq:constraint}) associated with solutions of the Maxwell system for static deformation of $\mathbb{S}^2$ subject to the initial conditions of Eq.~\eqref{eq:iniCondA} using adaptive RK$5(4)7$M as described in the text.
	Red: $L=8$, `$*$';
	Blue: $L=16$, `$\circ$';
	Green: $L=32$, `$\square$';
	Magenta: $L=64$, `$\lhd$'.
	Observe that constraints are well-preserved during the entire course of the solution - with the spectral (a) and pseudo-spectral (b) methods providing comparable accuracy. At $L\simeq 32$ we find that we have approximately saturated the convergence in $L$, that is, further increasing the band-limit will only marginally increase solution accuracy and a law of diminishing returns applies cf. Fig.~\ref{fig:nrg}.
}
\end{figure}

We now test the following initial conditions:
\begin{equation}
\label{eq:iniCondA}
\begin{aligned}
{}_0a_{2,0} &= 1;\\
{}_0\Phi_{1,1} &= \i; &{}_0\Phi_{1,-1} &= \i;\\
{}_0f_{0,0}&=10\sqrt{\pi}; &{}_0f_{2,0}&=1; &{}_0f_{4,3}&= 2\i; &{}_0f_{4,-3}&= 2\i.
\end{aligned}
\end{equation}
with all other values set to zero. This corresponds to a static deformation of $\mathbb{S}^2$, where at $t=0$ the deformation is chosen and fixed for all later times.
We calculate the solution numerically making use of the spectral and pseudo-spectral methods together with the embedded RK$5(4)7$M algorithm of \cite{DormPr:1980AFamRK}. This last choice of integrator allows for local error estimation and hence adaptive control of step-size in time.
In order to verify that our implementation is consistent we check
that the constraints are satisfied. 
This is done by (pseudo)spectral decomposition of Eq.~\eqref{eq:constraint} in
a similar manner to the preceding equations of this section. The results of
this are shown for both spectral (Fig.~\ref{fig:constr_sub1}) and
pseudo-spectral (Fig.~\ref{fig:constr_sub2}) implementations; we find excellent
agreement between both methods.

\begin{figure}[ht]
\centering
\begin{subfigure}[Spectral method]{%
    \includegraphics[width=.48\linewidth] {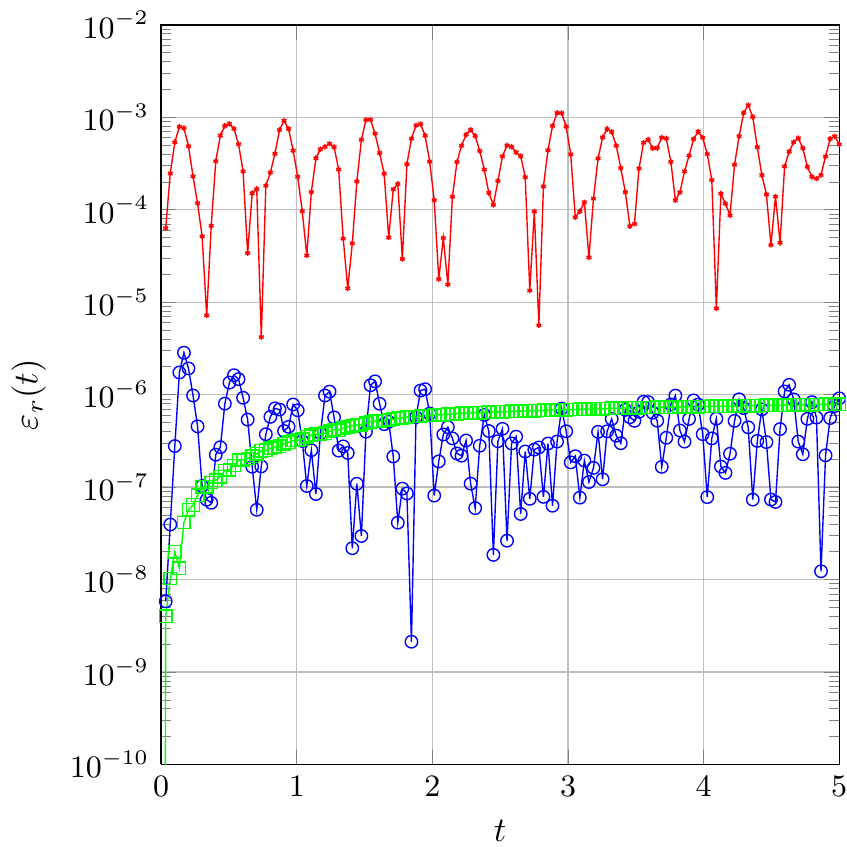}
    }
\end{subfigure}
\hfill
 \begin{subfigure}[Pseudo-spectral method]{%
     \includegraphics[width=.48\linewidth]{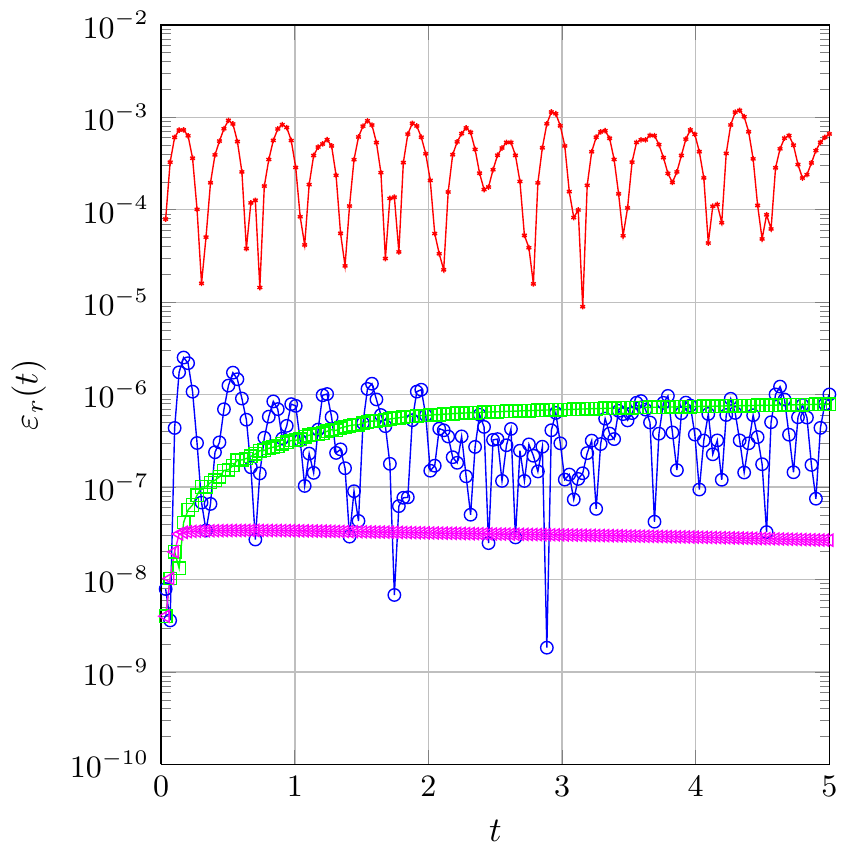}
     }
\end{subfigure}%
\caption{
\label{fig:nrg}
  (Colour online) The relative error in the energy $\varepsilon_r(t)$ as defined in Eq.~\eqref{eq:nrgRel} for numerical solution of the Maxwell system subject to the initial conditions of Eq.~\eqref{eq:iniCondA}.
 Red: $L=8$, `$*$';
 Blue: $L=16$, `$\circ$';
 Green: $L=32$, `$\square$';
 Magenta: $L=64$, `$\lhd$'.
 In both solutions, constructed via spectral (a) and pseudo-spectral (b)
 methods respectively, we find that oscillations in $\varepsilon_r(t)$ are
 observed for $L\in\{8,\,16\}$. We ascribe the aforementioned oscillations to
 the numerical scheme and not to any physical property of the Maxwell system as
 for $L\in\{32,\,64\}$ these spurious features disappear. It is important to
 note that even though oscillatory behaviour in $\varepsilon_r(t)$ is observed,
 it is bounded over the interval $t$ shown.
}
\end{figure}

Checking other invariants of the system can give further insight into the performance of our numerical scheme. Consider the energy as a function of time:
\begin{equation}\label{eq:nrgIntegral}
\mathcal{E}(t)=\frac{1}{8\pi}\int^{2\pi}_0\int_0^{\pi}\,\left(F_0F_0+2F_{-1}F_1\right)\frac{\sin\vartheta}{f^2}\,d\vartheta d\varphi.
\end{equation}
Due to Eq.~\eqref{eq:FreCond} we can conclude that Eq.~\eqref{eq:nrgIntegral}
is the integral of a positive definite quadratic form (since
$(\sin\vartheta)/f^2\geq0$ for $\vartheta\in[0,\pi]$) and hence
$\mathcal{E}(t)\geq0$ for all $t$. For time-independent $f$, $\partial_t$ is a
Killing vector and hence by Noether's theorem if $t_0$ is the initial time,
then we have $\mathcal{E}(t_0)=\mathcal{E}(t)$. That is, under a static deformation 
energy is conserved. In order to examine any deviations in $\mathcal{E}$ that may occur due to
numerical error, it is convenient to further define relative error in the
energy via:
\begin{equation}\label{eq:nrgRel}
\varepsilon_r(t):=\left|\frac{\mathcal{E}(t_0) - \mathcal{E}(t) }{\mathcal{E}(t)}\right|,
\end{equation}
which serves as a measure of the failure of energy conservation. We show the
value of $\varepsilon_r$ in Fig.~\ref{fig:nrg}
for the solution of the Maxwell
system with the initial conditions of Eq.~\eqref{eq:iniCondA}, again both
spectral and pseudo-spectral methods perform with consistent accuracy affirming
our intuition about energy conservation for the system in the case of static
deformation.

\begin{figure}[ht]
\centering
\begin{subfigure}[$t=0$ $\tilde{E}=1/100$;]{
    \includegraphics[width=.47\linewidth] {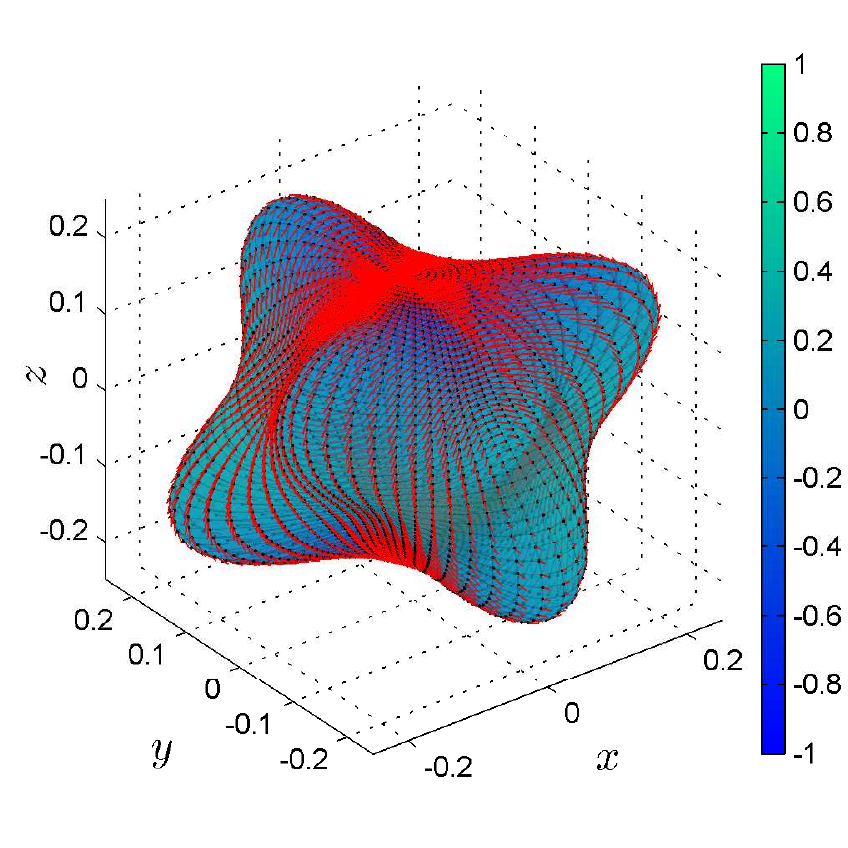}
    }
\end{subfigure}
\hfill
 \begin{subfigure}[$t=0.83$ $\tilde{E}=1/100$;]{
     \includegraphics[width=.47\linewidth]{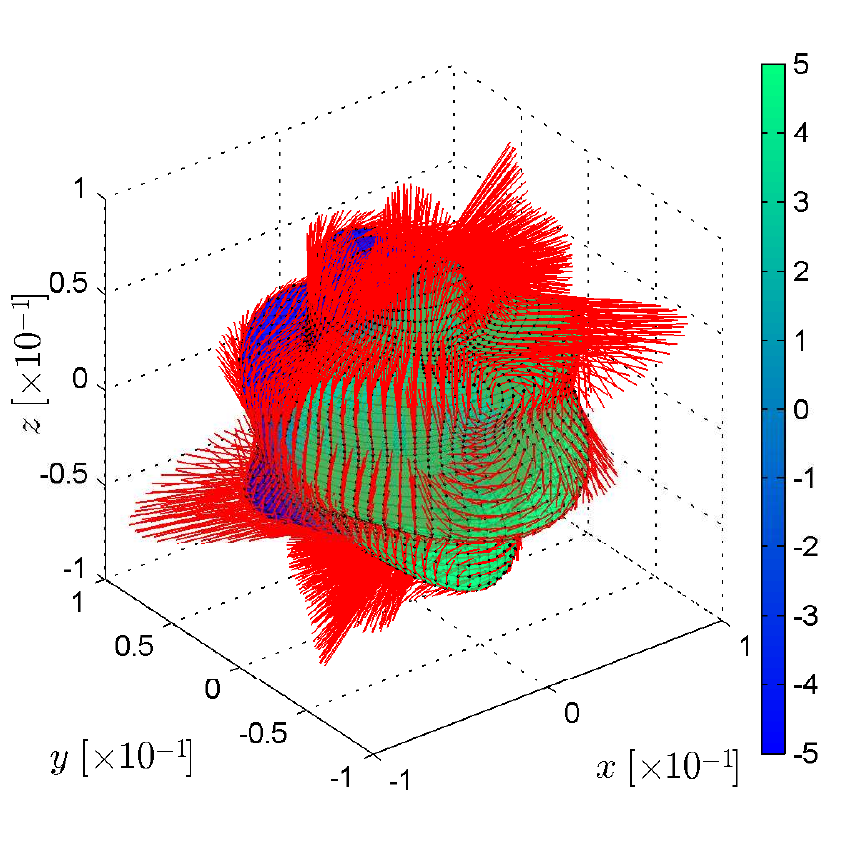}
     }
\end{subfigure}
\begin{subfigure}[$t=1.66$, $\tilde{E}=1/1600$;]{
    \includegraphics[width=.47\linewidth] {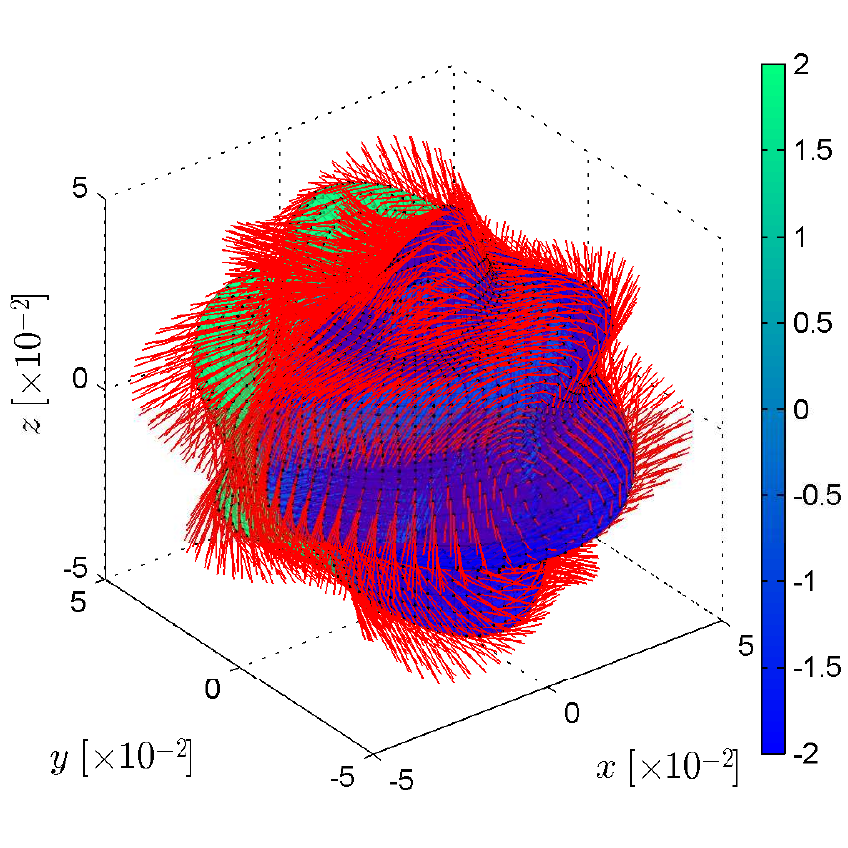}
    }
\end{subfigure}
\hfill
 \begin{subfigure}[$t=2.50$, $\tilde{E}=1/3200$;]{
     \includegraphics[width=.47\linewidth]{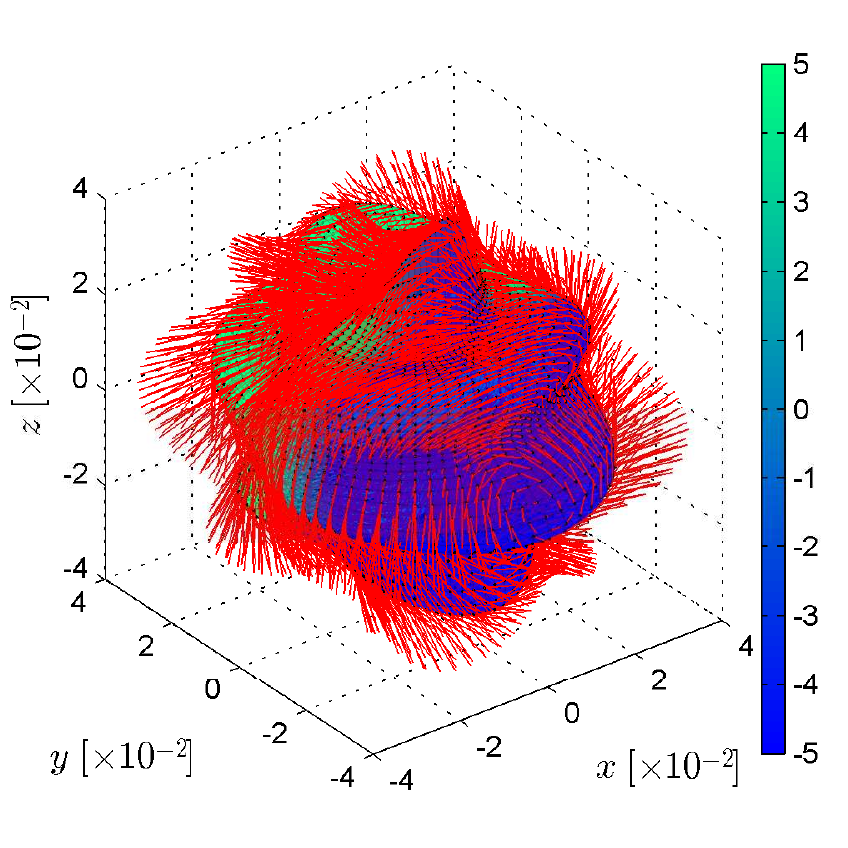}
     }
\end{subfigure}
\caption{
\label{fig:maxSoln}
(Colour online)
Maxwell $2+1$ solution, together with geometry as constructed by pseudo-spectral method at indicated times. Parameters: $L=32$, initial conditions and time-dependent $f$ as in Eq.~\eqref{eq:iniCondB} and Eq.~\eqref{eq:confTdep}. In each subfigure we show the electric field (red) scaled by the dimensionless factor $\tilde{E}$ together with the magnetic scalar. Nodes corresponding to the spatial sampling are indicated in black on the surface.
Note: Color and axial scalings are distinct over figures.
}
\end{figure}

\begin{figure}[ht]
\centering
\begin{subfigure}[Pseudo-spectral constraints]{%
    \includegraphics[width=.48\linewidth] {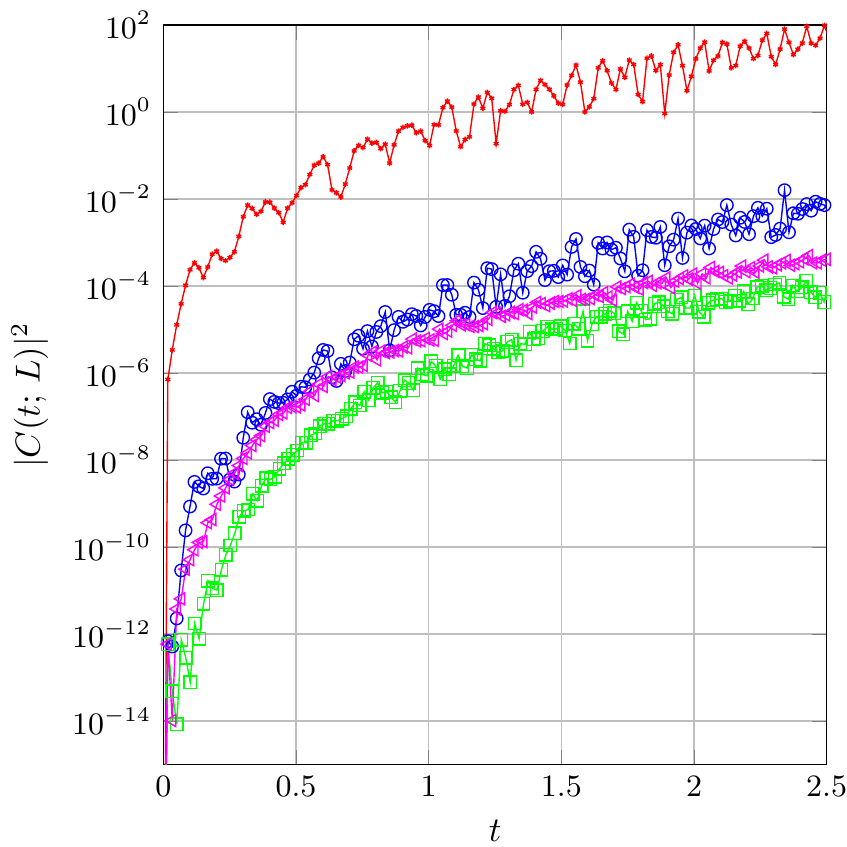}
   \label{fig:tdepconstr}}%
\end{subfigure}
\hfill
 \begin{subfigure}[Pseudo-spectral energy]{%
     \includegraphics[width=.48\linewidth]{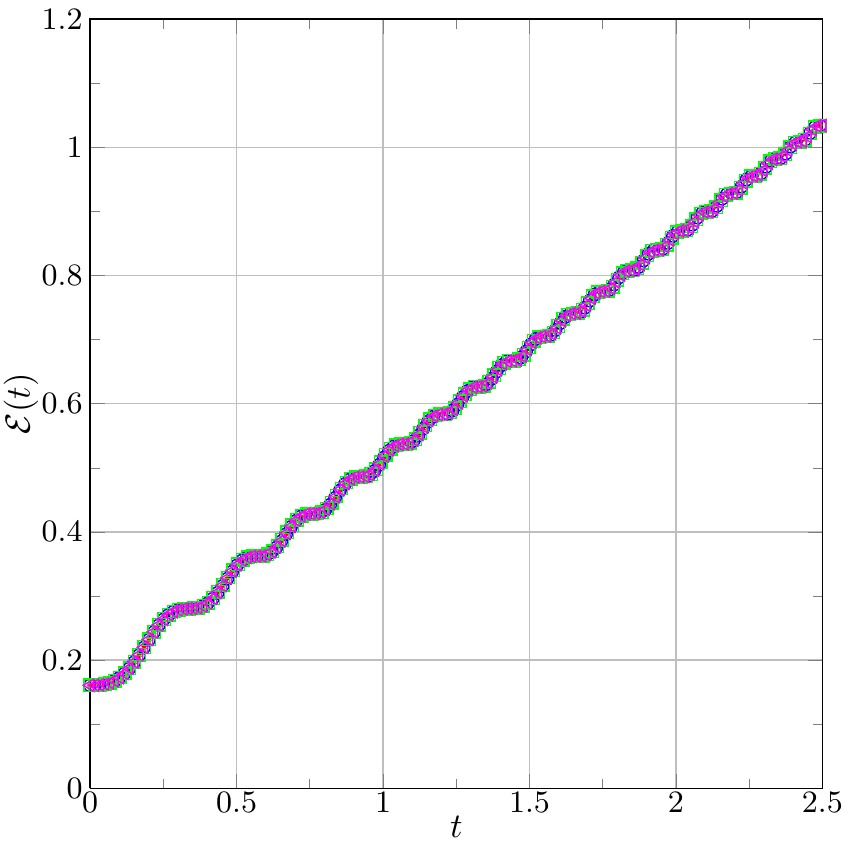}
     \label{fig:tdepnrg}}%
\end{subfigure}%
\label{fig:tdepconstrnrg}
\caption{
(Colour online)
Constraints and energy corresponding to the Maxwell-system with initial data prescribed by Eq.~\eqref{eq:iniCondB} and time-dependent spatial conformal factor of Eq.\eqref{eq:confTdep} as calculated by pseudo-spectral scheme.
 Red: $L=16$, `$*$';
 Blue: $L=32$, `$\circ$';
 Green: $L=64$, `$\square$';
 Magenta: $L=128$, `$\lhd$',
We find similar convergence behaviour of the contraints (a) to that of the time-independent deformation case, however we note that higher spatial resolution (band-limit $L$) is required. In (b) we find that the relation $\mathcal{E}(t)\geq 0$ is indeed satisfied.
}
\end{figure}

We now allow a time-dependent deformation of $\mathbb{S}^2$ using
the pseudo-spectral method. 
The direction $\partial_t$ is no longer a Killing
vector and energy expression of Eq.~\eqref{eq:nrgIntegral} will no longer be
conserved, however we still have $\mathcal{E}(t)\geq0$ for all $t$. For initial
conditions, we take:
\begin{equation}
\label{eq:iniCondB}
\begin{aligned}
{}_0a_{2,0} &= 1;\\
{}_0\Phi_{1,1} &= \i; &{}_0\Phi_{1,-1} &= \i;\\
{}_0g_{0,0}&=10\sqrt{\pi}; &{}_0g_{2,0}&=1; &{}_0g_{4,3}&= 2\i; &{}_0g_{4,-3}&= 2\i;\\
{}_0h_{0,0}&=12\sqrt{\pi}; &{}_0h_{2,0}&= 1; &{}_0h_{8,-1}&= 2\i; &{}_0h_{8,1}&= 2\i,
\end{aligned}
\end{equation}
where
\begin{align}\label{eq:confTdep}
{}_0f(t,\vartheta,\varphi)=&\frac{1}{t_c}\left(t_c-t\right){}_0g(\vartheta,\varphi)+t\,{}_0h(\vartheta,\varphi),&t_c:=2t_f,
\end{align}
governs the time-dependent deformation of $\mathbb{S}^2$ over an interval
$t\in[0,t_f]$. We show the solution at selected times, together with the
geometric picture in Fig.\ref{fig:maxSoln}.
As $f$ is now time-dependent the constraint propagation equation
\eqref{eq:constrEvF} suggests that numerical violations of the
constraints grow
with
time. 
Convergence of $C$ to a well-defined value with increasing band-limit $L$
is still expected and we find that a higher band-limit $L$ is required (see
Fig.\ref{fig:tdepconstr}) in constrast to the static (no deformation) case.
Finally we note that as anticipated, $\mathcal{E}(t)\geq0$ for $t\in[0,t_f]$ (see Fig.\ref{fig:tdepnrg}).

\section{Conclusion}
\label{sec:conclusion}

In this work we have presented a method for evolving tensorial equations on
manifolds with spherical topology. It is based on the use of the spin-weighted
spherical harmonics, a class of functions on the sphere which is closely
related to irreducible representations of $SU(2)$. We have demonstrated that
our method exhibits the accuracy and rapid convergence to solutions that is
expected from spectral methods.

Of course, this method is not limited to a single sphere. Instead, our main
application will be in systems defined on a manifold with spatial topology
$\mathbb{R}\times \St$, where the tensor fields are split into various pieces
intrinsic to the sphere factor and depending on a `radial' coordinate
corresponding to the $\mathbb{R}$ factor. This kind of topology occurs
naturally in the description of the global structure of space-times in general
relativity.

This method can be further generalised to half-integer spin. This will allow
us to solve spinorial equations, such as the Dirac equation or Weyl's equation
for a (massless) neutrino on space-times with spheroidal components as
discussed above.


 \section*{Acknowledgments}
 This research was partly funded by the Marsden Fund of the Royal Society of New
 Zealand under contract number UOO0922.


\end{document}